\documentclass[10pt,aps,twocolumn,preprintnumbers,prd,noshowpacs,nofootinbib,noshowkeys,floatfix,superscriptaddress]{revtex4-2}

\usepackage{amsmath,amssymb}
\usepackage{placeins}
\usepackage{comment}
\newcommand\redout{\bgroup\markoverwith
{\textcolor{red}{\rule[.5ex]{5pt}{0.7pt}}}\ULon}

\usepackage{graphicx}
\usepackage[colorlinks = true, 
            linktoc    = all, 
            linkcolor  = black, 
            citecolor  = blue, 
            urlcolor   = magenta]{hyperref} % Links in text
\usepackage{ulem}

\begin{document}
%\title{\texttt{BHAC-QGP}, a new code for (3+1)-dimensional simulations of relativistic heavy-ion collisions: application to Au-Au collisions}
\title{\texttt{BHAC-QGP}: three-dimensional MHD simulations of relativistic heavy-ion collisions \\ II. Application to Au-Au collisions}
\date{\today}
\author{Markus Mayer}
\email[E-mail: ]{mmayer@itp.uni-frankfurt.de}
\affiliation{Goethe University Frankfurt, Max-von-Laue-Straße 1, 60438 Frankfurt am Main, Germany}
\author{Ashutosh Dash}
\affiliation{Goethe University Frankfurt, Max-von-Laue-Straße 1, 60438 Frankfurt am Main, Germany}
\author{Gabriele Inghirami}
\affiliation{GSI Helmholtzzentrum für Schwerionenforschung, Darmstadt, Germany}
\author{Hannah Elfner}
\affiliation{GSI Helmholtzzentrum für Schwerionenforschung, Darmstadt, Germany}
\affiliation{Goethe University Frankfurt, Max-von-Laue-Straße 1, 60438 Frankfurt am Main, Germany}
\affiliation{Frankfurt Institute for Advanced Studies, Ruth-Moufang-Straße 1, 60438 Frankfurt am Main, Germany}
\affiliation{Helmholtz Research Academy Hesse for FAIR, Max-von-Laue-Straße 12, 60438 Frankfurt am Main, Germany}
\author{Luciano Rezzolla}
\affiliation{Goethe University Frankfurt, Max-von-Laue-Straße 1, 60438 Frankfurt am Main, Germany}
\affiliation{Frankfurt Institute for Advanced Studies, Ruth-Moufang-Straße 1, 60438 Frankfurt am Main, Germany}
\affiliation{School of Mathematics, Trinity College, Dublin 2, Ireland}
\author{Dirk H.\ Rischke}
\affiliation{Goethe University Frankfurt, Max-von-Laue-Straße 1, 60438 Frankfurt am Main, Germany}
\affiliation{Helmholtz Research Academy Hesse for FAIR, Max-von-Laue-Straße 12, 60438 Frankfurt am Main, Germany}

\begin{abstract}
We present \texttt{BHAC-QGP}, a new numerical code to simulate the evolution of matter created in heavy-ion collisions. \texttt{BHAC-QGP} is based on the Black Hole Accretion Code (\texttt{BHAC}), which has been designed to model astrophysical processes through the solution of the equations of general-relativistic magnetohydrodynamics. 
Like the mother code, \texttt{BHAC-QGP} uses Adaptive Mesh Refinement (AMR), which allows for a dynamic adjustment of the resolution in regions of the computational domain where a particularly high accuracy is needed.
We here discuss a number of applications of \texttt{BHAC-QGP} to Au-Au collisions at Relativistic Heavy-Ion Collider (RHIC) energies and show that the code is able to reproduce results of other simulations of these scenarios, but with much higher accuracy.

%\begin{description}
%\item[Background] This part would describe the context needed to understand what the paper is about.
%\item[Purpose] This part would state the purpose of the present paper.
%\item[Method] This part describe the methods used in the paper.
%\item[Results] This part would summarize the results.
%\item[Conclusions] This part would state the conclusions of the paper.
%\end{description}
\end{abstract}
\maketitle
%
% % % % % % % % % % % % % % % % % % % % % % % % % % % % % % % % % % % % % % % %
% BACKGROUND                                                                  %
% % % % % % % % % % % % % % % % % % % % % % % % % % % % % % % % % % % % % % % %
\section{Background}
In non-central relativistic heavy-ion collisions, the presence of charged protons leads to the generation of immensely strong electromagnetic fields \cite{Skokov:2009qp, Tuchin:2014hza, Tuchin:2013apa, Tuchin:2013ie, Bzdak:2011yy, Deng:2012pc, McLerran:2013hla, Voronyuk:2011jd, Li:2016tel, Gursoy:2014aka, Bzdak:2011yy, Pasechnik:2016wkt}. 
For instance the magnitude of the magnetic field in the laboratory frame can be estimated with the help of the Li\'{e}nard-Wiechert potentials as:
 \begin{equation}
   e B \approx \dfrac{e^2}{4 \pi}Z v \gamma \dfrac{8}{b^2}  \;. \label{Background1}
 \end{equation}
%
% Note: E-Field in co-moving frame from Coulomb law: E ~ Ze/R^2
% B-Field in lab frame from Boost: B ~ gamma*v*E ~ gamma v Ze/R^2 ~ gamma v Ze b/R^3 for b ~ R
%
Here, $Z$ represents the charge number and $R$ the radius of the nucleus, $e \approx 0.303$ denotes the electric charge, and $\gamma = \left(1-v^2\right){}^{-1/2}$ is the Lorentz factor of the nucleus in motion at velocity $v$.
This Lorentz factor is related to the center-of-mass energy $\sqrt{s_{\text{NN}}}$ via $\gamma = \sqrt{s_{\text{NN}}}/(2 m_N)$, with $m_N$ being the mass of the nucleon. 
Furthermore, $b$ in Eq.\ (\ref{Background1}) corresponds to the impact parameter, which is defined as the magnitude of the vector connecting the centers of the two colliding nuclei at the moment of their maximum geometrical overlap. 
Typically, the impact parameter establishes the direction of the $x$-axis. 
Together with the beam direction, which is usually taken along the $z$-axis, the impact parameter forms the so-called reaction plane. 
Notably, the generated magnetic field is oriented mainly in the direction perpendicular to the reaction plane. 
\par
The Relativistic Heavy-Ion Collider (RHIC) at the Brookhaven National Laboratory (BNL) collides Au nuclei with an energy of $\sqrt{s_{\text{NN}}} = 200~\text{GeV}$.
According to Eq.\ (\ref{Background1}), for an impact parameter of approximately $b \sim 10~\text{fm}$ such a collision produces a magnetic field of the order of $e B \approx 10~m_{\pi}^2 \approx 6 \times 10^{18}~\text{G}$, where $m_{\pi} \approx 0.14~\text{GeV}$ is the mass of a pion~\cite{Huang:2015oca}.
Hence, the magnetic field strength exceeds not only the squared pion mass but also the squared masses of up and down quarks. 
Such strong fields can induce substantial quantum effects in the produced quark-gluon plasma (QGP).
%
%(Note: $1 \text{MeV}^2 = e \cdot 1.6904 \cdot 10^{14} \text{Gauss} \cdot \hbar c^2$). [Thus, the magnetic field is larger than the squared masses of up or down quarks and thus is capable of inducing significant quantum effects.] 
%
Notably, due to the linear dependence on the center-of-mass energy, the magnetic fields generated in Pb-Pb collisions at the Large Hadron Collider (LHC) at CERN are even stronger. 
Moreover, the electric field generated by the relativistically fast moving charges is also enhanced by a Lorentz gamma factor and therefore equally strong as the magnetic field (\ref{Background1}).
These enormously strong electromagnetic fields can induce measurable effects on the particles within the QGP \cite{Roy:2017yvg, Inghirami:2019mkc, Dubla:2020bdz, ALICE:2019sgg, Oliva:2020doe, Sun:2021psy}.
\par
Given the extreme densities and temperatures within the QGP, individual quarks are no longer confined within hadrons and can move freely and interact with each other. 
In such conditions the chiral symmetry of Quantum Chromodynamics (QCD), which is spontaneously broken in the vacuum by a nonvanishing quark condensate, is restored.
Lattice-QCD calculations have shown that this chiral-symmetry restoration occurs at temperatures $T > 170~\mathrm{MeV}$~\cite{Borsanyi:2010bp, Bazavov:2011nk}.
Topological fluctuations can lead to a chiral imbalance in the medium~\cite{Kharzeev:1998kz}, which combined with strong external electromagnetic fields lead to a wide range of novel effects, including the Chiral Magnetic Effect (CME), Chiral Separation Effect (CSE), and Chiral Magnetic Wave (CMW) (see, for instance, Refs.\ \cite{Kharzeev:2007jp, Pu:2014fva, Kharzeev:2015znc, Huang:2015oca}).
The CME is caused by an imbalance of left- and right-handed charges and results in a vector charge current of oppositely polarized charges along the direction of the external magnetic field~\cite{Fukushima:2008xe, Sadofyev:2010pr}. 
The CSE exists in a medium with a nonzero vector charge, resulting in an axial-vector current along the direction of the magnetic field.
The interaction between CME and CSE causes the CMW~\cite{Kharzeev:2010gd}.
Another effect which occurs in a globally rotating system of chiral fermions (i.e., in non-central heavy-ion collisions) is the Chiral Vortical Effect (CVE), which is the generation of an axial-vector current along the rotation axis. 
Such an axial-vector current is converted into a spin polarization of quarks in the QGP, with fermions having their spins preferably aligned with the vorticity \cite{Kharzeev:2010gr, Jiang:2015cva, Son:2009tf, Asakawa:2010bu, Stephanov:2012ki}.
\par
The key to an experimental observation of the CME, CSE, and CMW lies in the strength of the electromagnetic fields. 
If these fields are too weak prior to the thermalization of the fireball, they may not lead to significant effects. 
In fact, the lifetime of the electromagnetic field represents a major uncertainty. 
In the simplest way it can be estimated as $t_B \sim R/(\gamma v)$~\cite{Deng:2012pc, Huang:2015oca}. 
Since the spectators quickly leave the collision zone, the remaining fireball plays an important role because its composition can significantly prolong the decay of the electromagnetic fields. 
If a charged QGP forms rapidly, it will respond to the fields and modify their evolution. 
This phenomenon is attributable to Faraday induction, where an external magnetic field induces an electric current within the medium, thereby generating a magnetic field that counteracts the decaying external field.
\par
In order to understand the dynamics of the electromagnetic fields in heavy-ion collisions and their effects on chiral phenomena, a comprehensive understanding of the entire evolution of the produced matter is essential.
Relativistic hydrodynamics serves as a successful framework for modeling the evolution of the QGP near thermodynamical equilibrium. 
Second-order dissipative relativistic hydrodynamics has proven to be capable of reproducing key observables, such as the transverse-momentum spectra and the elliptic flow.
These measurements suggest that the medium is strongly interacting, leading to a fast thermalization of the fireball and a strongly interacting nature~\cite{Dumitru:1998es, Molnar:2004yh, Petersen:2015rra, Petersen:2014yqa, Heinz:2013th}.
Incorporating electromagnetic fields into hydrodynamics leads to a theory of magnetohydrodynamics (MHD).
\par
With the help of numerical MHD simulations, it is possible to calculate the (3+1)-dimensional evolution of the fireball under the influence of these electromagnetic fields, enabling the exploration of uncertainties and chiral effects. 
This paper presents the results of numerical simulations of heavy-ion collisions performed with \texttt{BHAC-QGP}, a new code capable of describing the space-time evolution of the QGP under the influence of magnetic fields. 
\par
This paper is organized as follows: We begin by providing a brief summary of the numerical formalism of \texttt{BHAC-QGP}. 
Next, we present the procedures implemented for initializing heavy-ion collisions within \texttt{BHAC-QGP}, specifically explaining the optical Glauber model and the equations used for calculating initial magnetic fields. 
Afterwards, we explain the application of the Cooper-Frye prescription, which we employ to evaluate the freeze-out of particles from the fluid evolution.
From the Cooper-Frye prescription, one can then compute particle spectra and anisotropic flow coefficients.
Following this introduction, we present the results derived from simulations performed with \texttt{BHAC-QGP}.
\par
Throughout this article we use natural Heaviside-Lorentz units, $\hbar = c = k_B = \epsilon_0 = \mu_0 = 1$. 
In these units the electric charge is dimensionless $e := \sqrt{4 \pi \alpha \hbar c} \approx 0.303$, where $\alpha \approx 1/137$ is the fine-structure constant. 
As signature of the metric tensor, we choose $\left(-,+,+,+\right)$. 
We use Greek indices to indicate the components of a four-vector, while Latin indices range from 1 to 3 for its spatial components. 
Bold letters indicate three-vectors. 
The scalar and vector products between three-vectors are as usual denoted by $\cdot$ and $\times$, respectively.
In this article we will work with two coordinate systems, Cartesian coordinates $(t,x,y,z)$ and Milne coordinates $(\tau,x,y,\eta_{S})$, where $\tau$ is the proper time and $\eta_S$ the so-called space-time rapidity.
%
% % % % % % % % % % % % % % % % % % % % % % % % % % % % % % % % % % % % % % % %
% FORMALISM                                                                  %
% % % % % % % % % % % % % % % % % % % % % % % % % % % % % % % % % % % % % % % %
\section{Formalism}
\begin{figure}[!ht]
   \centering
   \includegraphics[width=0.95\linewidth]{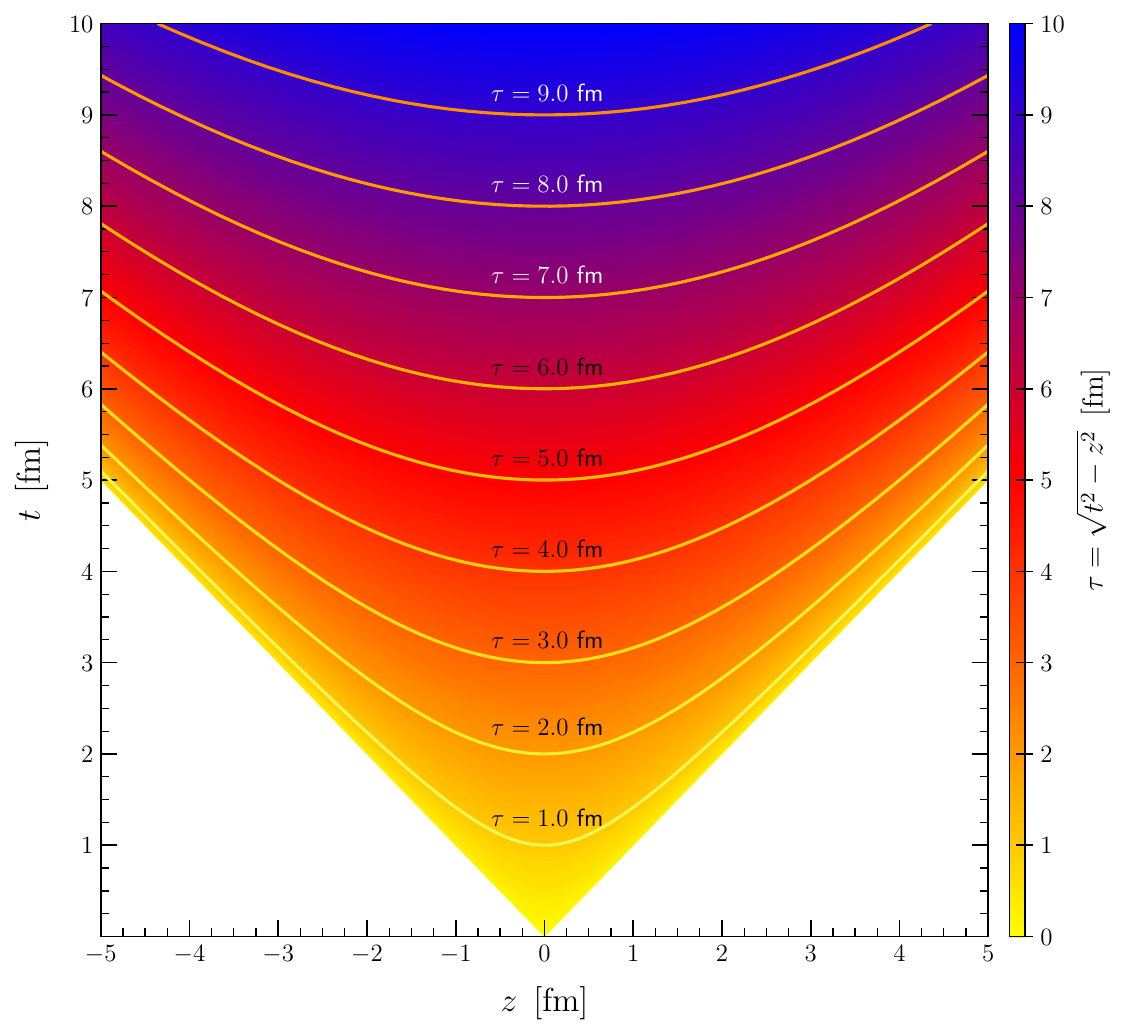}
      \caption{(1+1)-dimensional Milne diagram of a relativistic heavy-ion collision. 
   At $t = 0~\text{fm}$ the two ions collide at $z = 0~\text{fm}$ and produce an extremely hot fluid in the collision region. 
   This region approaches thermodynamical equilibrium after $\tau_0 \sim 1~\text{fm}$, so that it can be described with hydrodynamics. 
   This description is possible up to a certain freeze-out time, $\tau_{\text{f.o.}} \sim 10~\text{fm}$, after which the fluid becomes a gas of free-streaming particles.}
   \label{Fig:MilneCoordinates1}
 \end{figure}

This section is dedicated to our numerical setup of heavy-ion collisions. 
Our goal is to simulate the evolution of the matter produced in relativistic heavy-ion collisions under the influence of external magnetic fields. 
\texttt{BHAC-QGP} is capable to solve the equations of ideal general-relativistic MHD with any time-independent metric. 
For the purpose of calculating heavy-ion collisions, we implemented time-dependent Milne coordinates, because these are best suited to describe the longitudinal expansion of a fluid, cf.\ Fig.\ \ref{Fig:MilneCoordinates1}. 
\texttt{BHAC-QGP} benefits from the underlying parallel Adaptive Mesh Refinement (AMR) approach, that  allows for a dynamic adjustment of the numerical resolution. 
Because the magnetic fields in heavy-ion collisions also exist outside the collision zone, \texttt{BHAC-QGP} has to deal with regions where the magnetic pressure is much stronger than the fluid pressure. 
One-dimensional (1D) and two-dimensional (2D) inversion schemes are likely to fail in such cases. 
Fur such scenarios, \texttt{BHAC-QGP} is equipped with a so-called ``entropy switch'', such that in highly magnetized regions, \texttt{BHAC-QGP} can use the entropy evolution equation to find a valid solution. 
More details on the set of equations solved and on the numerical techniques employed for their solution can be found in a companion paper, hereafter referred to as Paper I. 
%\textcolor{red}{(DHR: need to insert a ref.!)}
%\textbf{[TODO: Link/REF to Paper 1 - Numerical schemes]}
%\FloatBarrier % Fix/Keep figures in the appropriate subsection
%
% % % % % % % % % % % % % % % % % % % % % % % % % % % % % % % % % % % % % % % %
% GLAUBER MODEL                                                               %
% % % % % % % % % % % % % % % % % % % % % % % % % % % % % % % % % % % % % % % %
 \subsection{Initial energy density - Glauber model}
The initial energy-density distribution of the matter produced in heavy-ion collisions is computed using the optical Glauber model \cite{Vogt:2007zz, Jaiswal:2016hex, dEnterria:2020dwq, Kolb:2001qz, DelZanna:2013eua, Karpenko:2013wva, Schenke:2010nt, Miller:2007ri, Florkowski:2010zz}, which is also used in many other codes (see, for example, Refs.\ \cite{DelZanna:2013eua, Karpenko:2013wva, Schenke:2010nt, Miller:2007ri, Vogt:2007zz, Florkowski:2010zz}). 
The basis of this model are two ions, projectile $A$ and target $B$, colliding at relativistic velocities and arbitrary impact parameter. 
The Glauber model establishes a relationship between the impact parameter $b$, the number of participants $N_{\text{part}}$, and the number of collisions $N_{\text{coll}}$. 
A small impact parameter corresponds to a \textit{central} collision, while collisions with a large impact parameter are called \textit{non-central} or \textit{peripheral}. 
\par
For estimating of $N_{\text{part}}$ and $N_{\text{coll}}$, the model requires only two ingredients: an inelastic nucleon-nucleon cross section $\sigma_{\text{inel}}$ and an assumption on the shape of the colliding nuclei. 
The inelastic cross section is usually taken from experimental measurements, while for the nucleus one commonly assumes a smooth mass density described by a Woods-Saxon distribution (see, for instance, Ref.\ \cite{Miller:2007ri}), 
 \begin{equation}
   \rho(x,y,z) := \dfrac{\rho_0}{1 + \text{exp}\!\left(\frac{\sqrt{x^2 + y^2 + z^2}-R}{d}\right)}  \;, \label{GlauberEq1}
 \end{equation}
where $R$ is the radius of the nucleus, which is related to the mass number $A$ via $R \approx (1.12 \, A^{1/3} - 0.86 \, A^{-1/3})~\text{fm}$~\cite{bohr:1998nsv}, while $d$ is the surface thickness of the nucleus. 
The parameter $\rho_0$ represents a normalization factor defined to ensure $\int \mathrm{d}^3 r\,\rho\!\left(r\right)  = A$. 
With the Woods-Saxon distribution\ (\ref{GlauberEq1}) it is then possible to evaluate the so-called nuclear profile function, which corresponds to the probability per unit transverse area of a nucleon being located in a projectile or target flux tube,
 \begin{equation}
   \hat{T}_{\text{A,B}}\!\left(x,y\right) := 
   \int\limits_{-\infty}^{\infty} \mathrm{d}z\, \hat{\rho}_{\text{A,B}}\!\left(x,y,z\right)   \;. \label{GlauberEq2}
 \end{equation}
Here, $\hat{\rho}_{\text{A,B}}$ is the Woods-Saxon distribution of either projectile (the mass number of which we continue to denote by $A$) or target (the mass number of which we denote by $B$ in the following), normalized to the total number of projectile and target particles, i.e, divided by $A$ for the projectile and $B$ for the target. 
Because the impact-parameter vector $\boldsymbol{b}$ defines the $x$-axis, $\boldsymbol{b} := \left(b, 0\right)^T$, the nuclear profile function actually depends only on the magnitude of the impact parameter, $\hat{T}_{\text{A,B}}\!\left(x,y\right) = \hat{T}_{\text{A,B}}\!\left(b\right)$. 
The integration over the joint probability then defines the so-called nuclear thickness function,
 \begin{equation}
   \hat{T}_{\text{AB}} := 
     \int\limits_{-\infty}^{\infty} \! \int\limits_{-\infty}^{\infty} \mathrm{d}x \, \mathrm{d}y\,
     \hat{T}_{\text{A}}\!\left(x+\dfrac{b}{2},y\right) 
     \hat{T}_{\text{B}}\!\left(x-\dfrac{b}{2}, y\right)    \;. \label{GlauberEq3}
 \end{equation}
The nuclear thickness function $\hat{T}_{\text{AB}}$ can be interpreted as effective overlap area for which a specific nucleon in projectile $A$ interacts with a given nucleon in target $B$. 
The probability of a collision between two particles in the transverse plane, that is the number density of collisions, is then simply the product of nuclear thickness function and inelastic nucleon-nucleon cross section,
 \begin{equation}
   n_{\text{coll}} = \sigma_{\text{inel}} \, \hat{T}_{\text{AB}}  \;,  \label{GlauberEq4}
 \end{equation}
so that the total number of collisions has a strong dependence on the impact parameter. 
In addition to $n_{\text{coll}}$, the number density of participating nucleons $n_{\text{part}}$ can also be estimated as:
 \begin{align}
   \lefteqn{n_{\text{A}; \text{part}}\left(x,y;b\right) =} \notag \\
   &  A \, \hat{T}_{\text{A}}\!\left(x + \frac{b}{2}, y\right) \left\{1 - \left[1 - \hat{T}_{\text{B}}\!\left(x - \frac{b}{2}, y\right) \sigma_{\text{inel}}\right]^B\right\}  \;, \label{GlauberEq5}   \\
   \lefteqn{n_{\text{B}; \text{part}}\left(x,y;b\right) =} \notag \\
   &  B \, \hat{T}_{\text{B}}\!\left(x - \frac{b}{2}, y\right) \left\{1 - \left[1 - \hat{T}_{\text{A}}\!\left(x + \frac{b}{2}, y\right) \sigma_{\text{inel}}\right]^A\right\}  \;. \label{GlauberEq6} 
 \end{align}
The total density of participants is then $n_{\text{part}}\ = n_{\text{A}; \text{part}} + n_{\text{B}; \text{part}}$. 
Since experimental measurements indicate that the deposited energy density scales with the number density of participants~\cite{Kharzeev:2000ph, Hirano:2001eu}, we follow Refs.\ \cite{DelZanna:2013eua, Schenke:2010nt} and parametrize the initial energy density in the transverse plane (at $z=\eta_S=0$) according to the two-component model as:
% \begin{eqnarray}
%   e_{\text{2D}}\!\left(x, y; b\right) &=& e_0 \Bigg[\alpha_H \dfrac{n_{\text{coll}}\!\left(x, y; b\right)}{n_{\text{coll}}\!\left(0, 0; 0\right)} \notag \\
%   &\qquad& + \left(1 - \alpha_H\right) \dfrac{n_{\text{part}}\!\left(x, y; b\right)}{n_{\text{part}}\!\left(0, 0; 0\right)}\Bigg] \label{GlauberEq7} 
% \end{eqnarray}
 \begin{eqnarray}
   \lefteqn{ e_{\mathrm{2D}}\!\left(x, y; b\right)  } \notag \\
   &=& e_0 \Bigg[\dfrac{\alpha_H n_{\text{coll}}\!\left(x, y; b\right) + \left(1 - \alpha_H\right) n_{\text{part}}\!\left(x, y; b\right)}{\alpha_H n_{\text{coll}}\!\left(0, 0; 0\right) + \left(1 - \alpha_H\right) n_{\text{part}}\!\left(0, 0; 0\right)}\Bigg]  \;, \label{GlauberEq7} 
 \end{eqnarray}
where $e_0 := e\!\left(0,0,0\right)$, i.e., the energy density $e$ at the origin for $b = 0$.
The parameter $\alpha_H \in \left[0,1\right]$ is the so-called collision hardness and allows for a scaling between \textit{hard} and \textit{soft} processes. 
Hard processes are high-energy interactions between quarks and gluons that produce particles such as jets, leptons, or pions with large transverse momenta, $p_T > 1.0~\text{GeV}$, which are neglected when $\alpha_H = 0$. 
Soft processes are characterized by longer time scales and smaller momentum transfers, which occur, for example, in hadronization or parton energy loss. 
Experimental measurements indicate that soft processes dominate in heavy-ion collisions \cite{Kharzeev:2000ph}. 
This means that most of the interactions involve little transverse-momentum transfer, such that $\alpha_H$ is usually relatively small. 
In summary, $e_0$ and $\alpha_H$ are the two tunable parameters of the two-dimensional Glauber model (\ref{GlauberEq7}), which can be fixed by comparing the particle distributions from the hydrodynamical model with experimental data. 
\par
\begin{figure}[!ht]
   \centering
   \includegraphics[width=1.00\linewidth]{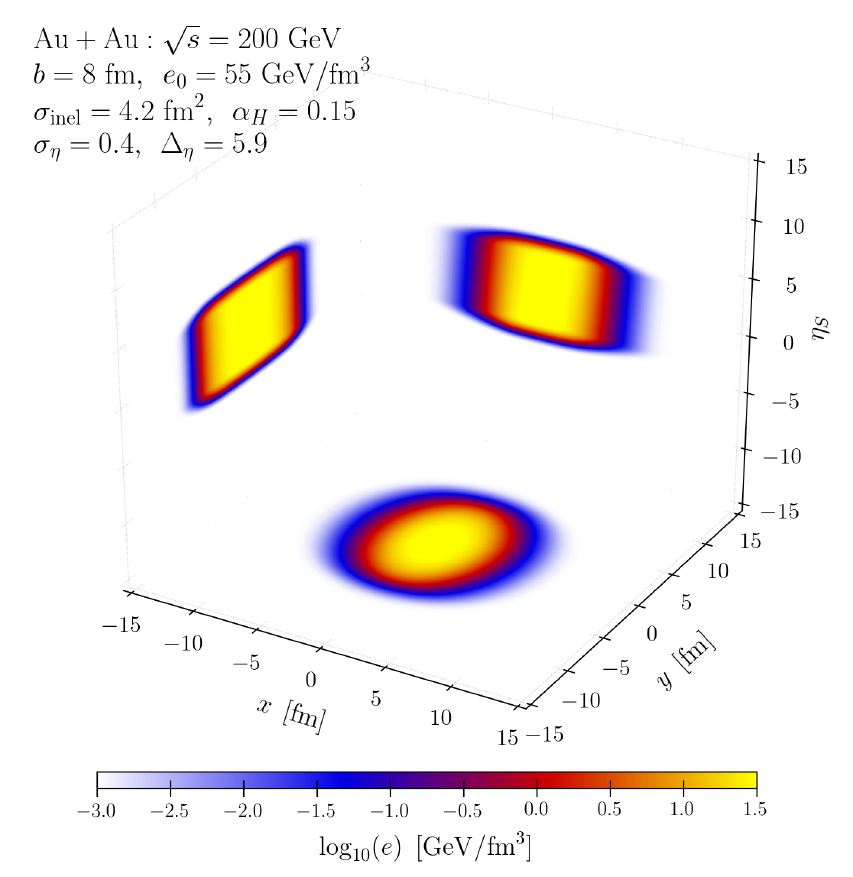}
   \caption{Initial energy-density distribution for a Au-Au collision at RHIC energy $\sqrt{s_{\text{NN}}} = 200~\text{GeV}$ at an impact parameter of $b = 8~\text{fm}$, calculated with the (3+1)-dimensional optical Glauber model. 
    }
   \label{Fig:InitEnergyDensity1}
 \end{figure}

Three-dimensional simulations of heavy-ion collisions have to take into account also the longitudinal dependence of the initial energy density. 
A widely used parameterization \cite{Schenke:2010nt, DelZanna:2013eua, Hirano:2001eu, Hirano:2002ds, Nonaka:2006yn} is based on a flat rapidity region around $\eta_S = 0$ and a Gaussian damping in forward and backward direction, 
 \begin{equation}
 \!\!
   H\!\left(\eta_S\right) = 
     \text{exp}\!\left[- \Theta\!\left(\vert \eta_S \vert - \frac{\Delta_{\eta_S}}{2}\right)
     \dfrac{\left(\vert \eta_S \vert - \Delta_{\eta_S}/2\right)^2}{2 \sigma_{\eta_S}^2}\right] . \label{GlauberEq8}
 \end{equation}
where the tunable parameter $\Delta_{\eta_S}$ controls the width of the flat rapidity region, while the parameter $\sigma_{\eta_S}$ sets the damping of the Gaussian fall-off. 
The full parameterization of the energy density reads then as:
 \begin{eqnarray}
   &e_{\text{3D}}&\left(x, y, \eta_S; b\right) = \dfrac{e_0}{w} \, H\!\left(\eta_S\right) \, \Theta\!\left(Y_B - \vert \eta_S \vert\right) \label{GlauberEq9} \\
   &\times& 
   \Bigg[\alpha_H n_{\text{coll}}\!\left(x, y; b\right) + \left(1 - \alpha_H\right) \notag \\
   && \times
   \Bigg(\dfrac{Y_B - \eta_S}{Y_B} n^{\text{A}}_{\text{part}}\!\left(x, y; b\right) + \dfrac{Y_B + \eta_S}{Y_B} n^{\text{B}}_{\text{part}}\!\left(x, y; b\right)\Bigg)\Bigg] \;.\notag
 \end{eqnarray}
The Heaviside function guarantees that initially the matter does not move faster in longitudinal direction than with beam rapidity,
% NOTE
%A longitudinal expansion of a fluid element which is initially at rest at $\eta_S = Y_B$ produces particles which move faster than $Y_B$)}, 
i.e., $\vert \eta_S \vert \le Y_B$, where the beam rapidity is defined as:
 \begin{equation}
   %Y_B := \text{ln}\!\left(\dfrac{\sqrt{s_{\text{NN}}}}{m_p}\right)
   Y_B := \dfrac{1}{2} \, \text{ln}\!\left(\dfrac{1 + v}{1 - v}\right)  \;. \label{GlauberEq10}
 \end{equation}
%where $m_p$ is the proton mass. 
The energy density in Eq.\ (\ref{GlauberEq9}) is weighted by a factor
 \begin{equation}
   w = \left(1-\alpha_H\right) n_{\text{part}}\!\left(0, 0; 0\right) + \alpha_H n_{\text{coll}}\!\left(0, 0; 0\right)  \;. \label{GlauberEq11}
 \end{equation}
 The initial energy-density distribution for a Au-Au collision at RHIC energy with impact parameter $b=8$ fm is shown in Fig.\ \ref{Fig:InitEnergyDensity1}.
 For such a peripheral collision the fireball has an almond-like shape in the $x$-$y$-plane (at $\eta_S = 0$). 
  \\
%\FloatBarrier % Fix/Keep figures in the appropriate subsection
%
% % % % % % % % % % % % % % % % % % % % % % % % % % % % % % % % % % % % % % % %
% MAGNETIC FIELD                                                              %
% % % % % % % % % % % % % % % % % % % % % % % % % % % % % % % % % % % % % % % %
 \subsection{Initial magnetic field}
Since protons are positively charged, they generate an electric current when in motion, which subsequently produces a magnetic field according to Maxwell's equations\cite{Li:2016tel, Inghirami:2019mkc,Dash:2023kvr}. 
Because heavy-ion collisions take place at almost the speed of light, the resulting magnetic fields also reach enormous strengths. 
Sufficiently strong magnetic fields may have consequences on the dynamics of the produced QGP and on the particles itself. 
Therefore, to ultimately study the effect of the electromagnetic fields, a realistic initial parametrization of these fields produced during the collision of the two positively charged ions is required. 
Earliest calculations indicated that the electromagnetic field could decay and even disappear so quickly that it may not have a considerable effect on the QGP (see, e.g., Ref.\ \cite{Oliva:2020mfr}). 
However, the situation is quite different when the medium is electrically conducting. 
Indeed, a QGP with a large electric conductivity will have a significant response to the change of the external magnetic field, delaying the decay of the electromagnetic field. 
This is simply a consequence of the induced currents: An external magnetic field will induce an electric current in the medium which, in turn, will cause a magnetic field compensating the decaying external magnetic field. 
Lattice-QCD calculations predict that the electric conductivity of a static QGP is relatively large~\cite{Ding:2010ga, Amato:2013naa, Aarts:2014nba, Li:2016tel, Pasechnik:2016wkt},
 \begin{equation}
   \sigma_E = \left(5.8 \pm 2.9\right) \dfrac{T}{T_C} ~\text{MeV}\;,
 \end{equation}
where $T$ is the plasma temperature and $T_C$ the transition temperature between QGP and hadronic phase. 
However, these calculations are done in the quenched approximation (without dynamical quarks) and assume that the medium itself is static. 
Furthermore, a uniform but temperature-dependent electric conductivity is clearly an approximation since the temperature in a collision varies rapidly in spacetime.
\par
\begin{figure*}[!htp]
    \centering
    \includegraphics[width=0.33\linewidth]{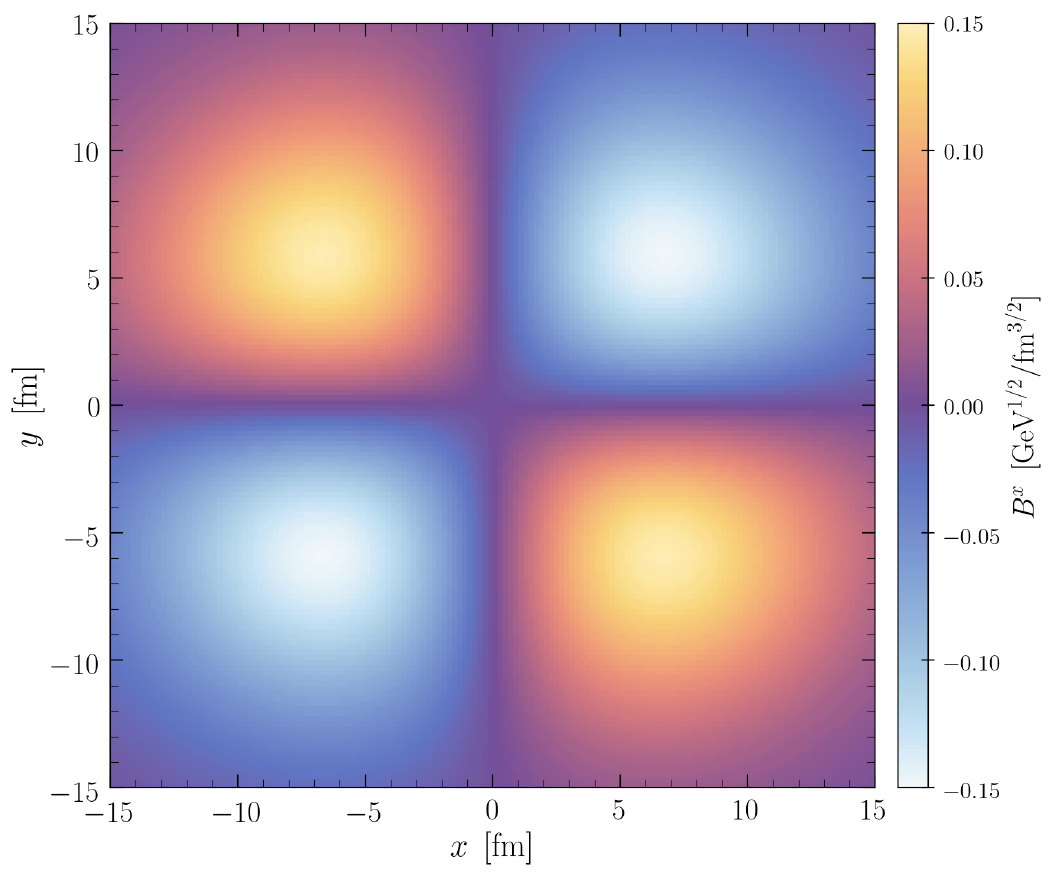}\hfill
    \includegraphics[width=0.33\linewidth]{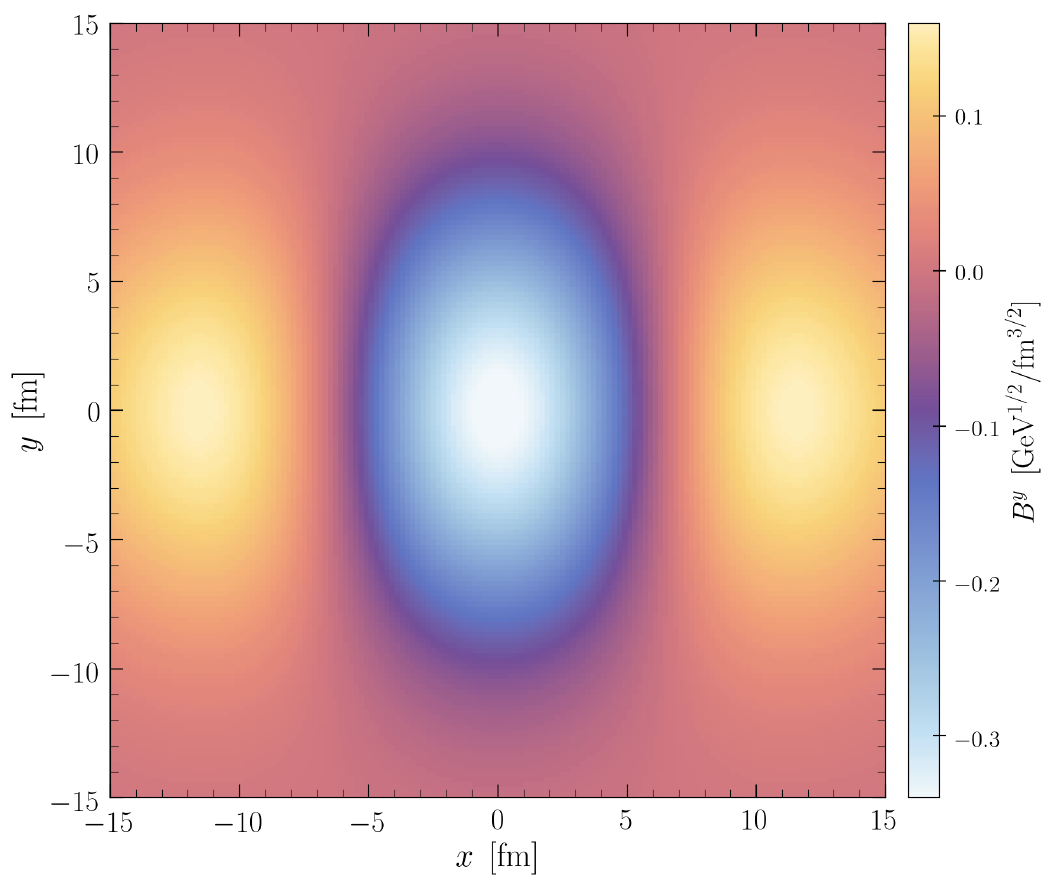}\hfill
    \includegraphics[width=0.33\linewidth]{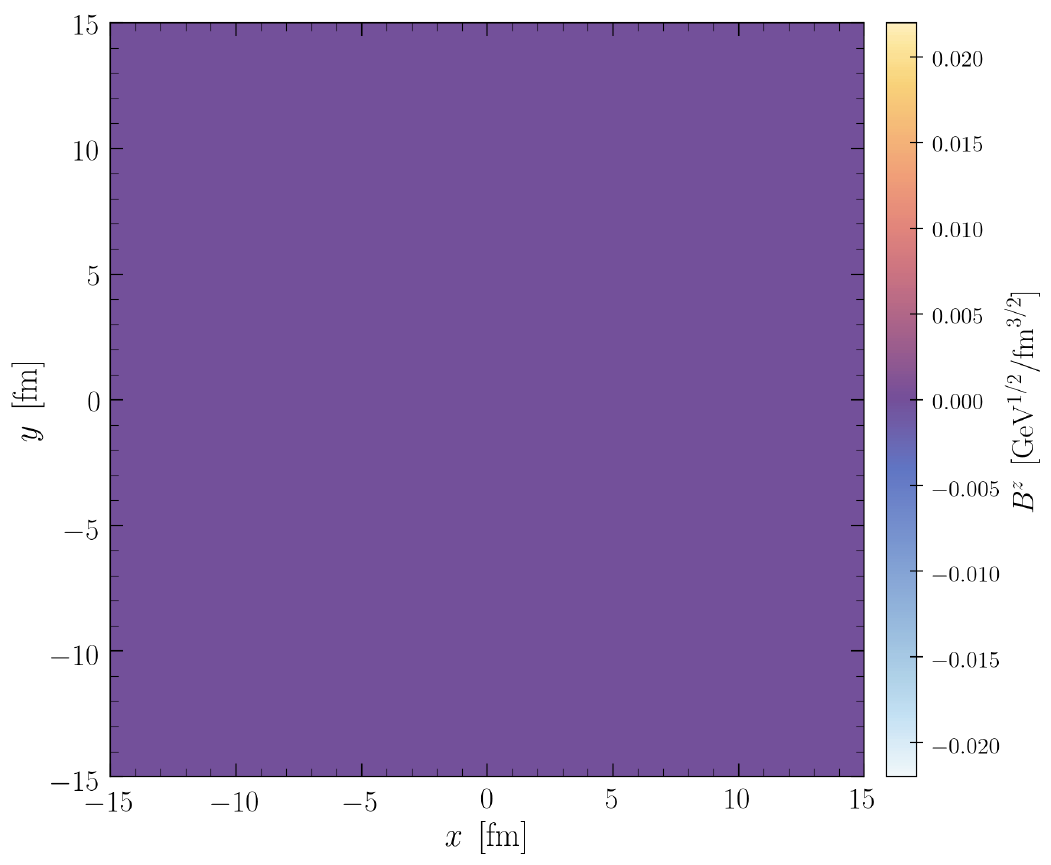}\\
    \includegraphics[width=0.33\linewidth]{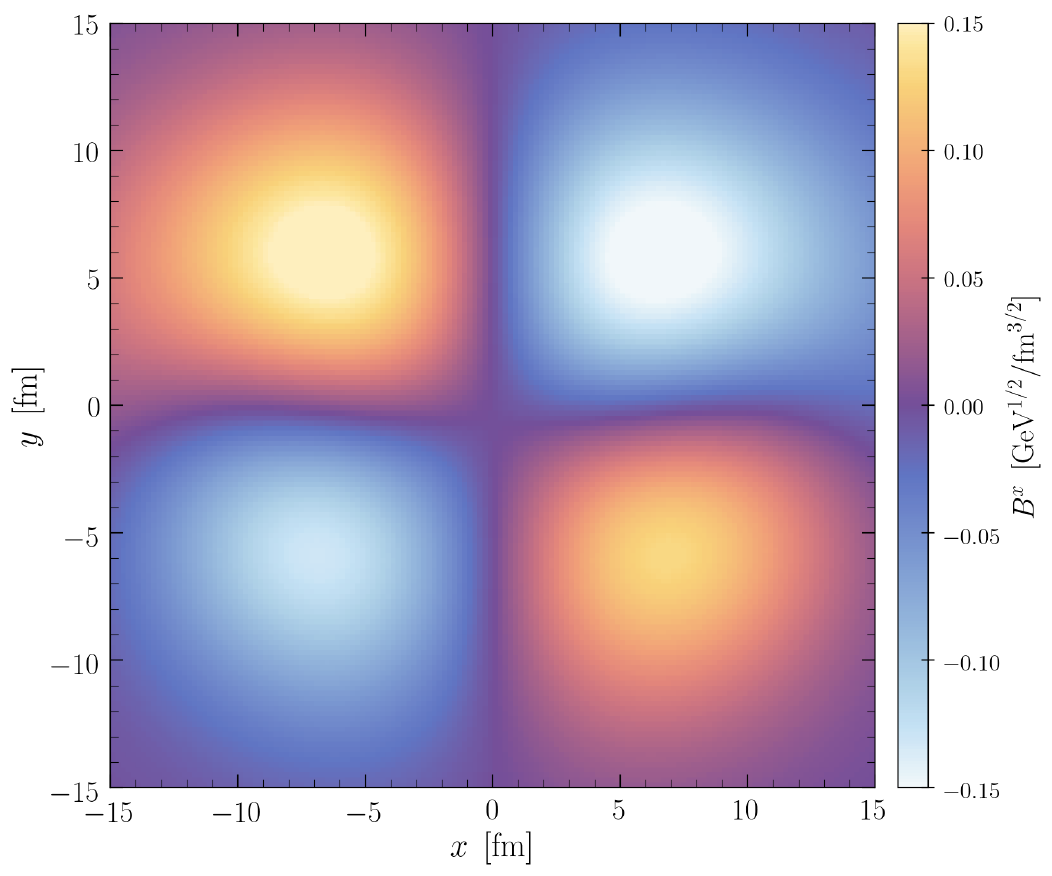}\hfill
    \includegraphics[width=0.33\linewidth]{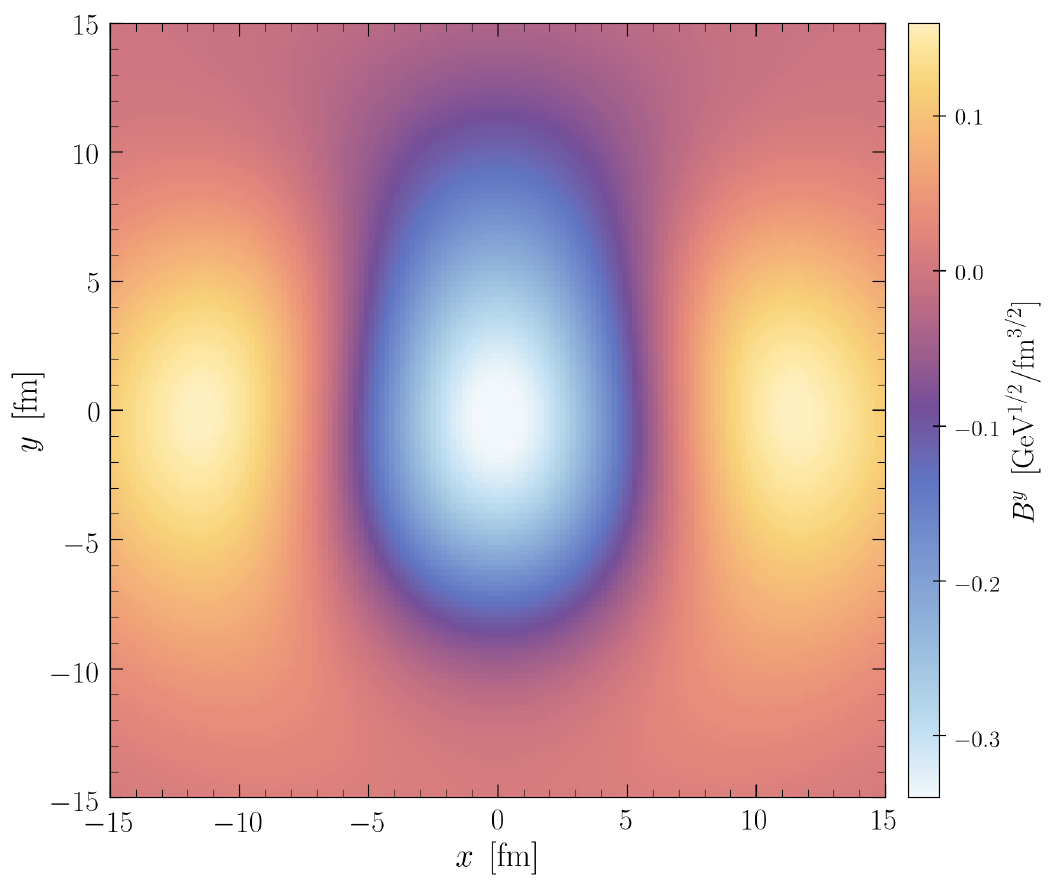}\hfill
    \includegraphics[width=0.33\linewidth]{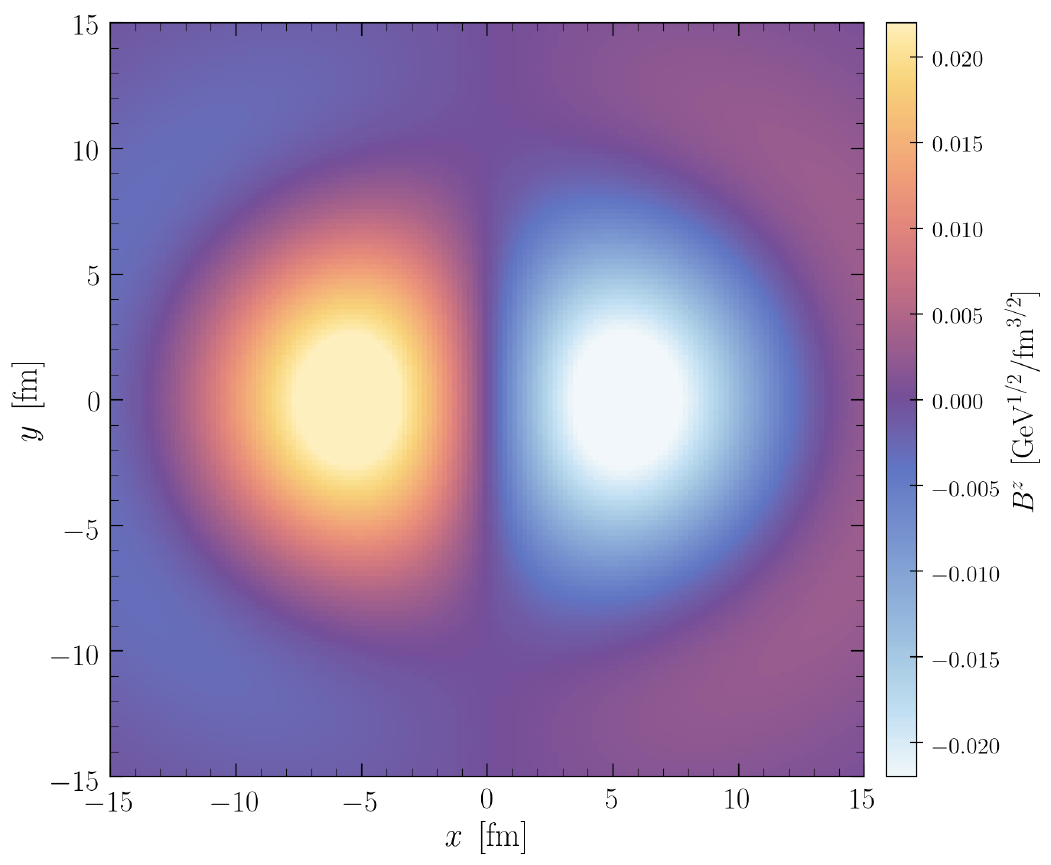}
    \caption{Initial magnetic field in the transverse plane (at $\eta_S = 0$) as it is produced in a Au-Au collision at a center-of-mass energy $\sqrt{s_{\text{NN}}} = 200~\text{GeV}$ and an impact parameter of $b = 10~\text{fm}$. 
    The top row shows the magnetic field for a medium with $\sigma_E = 5.80~\text{MeV}$, $\sigma_{\chi} = 0~\text{MeV}$, while the bottom row represents the magnetic field calculated for a medium with $\sigma_E = 5.80~\text{MeV}$, $\sigma_{\chi} = 1.50~\text{MeV}$. 
    }
    \label{Fig:InitMagneticField1}
  \end{figure*}

As described in Ref.\ \cite{Li:2016tel}, it is possible to solve Maxwell's equations for an infinite homogeneous medium that has a constant electric conductivity $\sigma_E$ and a constant chiral magnetic conductivity $\sigma_{\chi} \ll \sigma_E$. 
The magnetic field created by one particle with electrical charge $e$ moving in $z$-direction with velocity $v$ is a solution of the Maxwell equations, which in cylindrical coordinates takes the form:
 \begin{align}
   B_{\rho}\!\left(t, \boldsymbol{x}\right) &= -\sigma_{\chi} \dfrac{e}{8 \pi} \dfrac{v \gamma^2 x_T}{\Delta^{3/2}} \left[\gamma \left(v t - z\right) + \varkappa \sqrt{\Delta}\right] e^\varkappa  \;, \\
   B_{\phi}\!\left(t, \boldsymbol{x}\right) &= \dfrac{e}{4 \pi} \dfrac{v \gamma x_T}{\Delta^{3/2}} \left(1 + \dfrac{\sigma_E v \gamma}{2} \sqrt{\Delta}\right) e^\varkappa  \;, \\
   B_{z}\!\left(t, \boldsymbol{x}\right) &= \sigma_{\chi} \dfrac{e}{8 \pi} \dfrac{v \gamma}{\Delta^{3/2}} \Big[\gamma^2 \left(vt - z\right)^2 
 \\ &  \times
 \left(1 + \dfrac{\sigma_E v \gamma}{2} \sqrt{\Delta}\right) + \Delta \left(1 - \dfrac{\sigma_E v \gamma}{2} \sqrt{\Delta}\right)\Big] e^\varkappa  \;, \notag 
 \end{align}
where $x_T^2 := x^2 + y^2$, $\Delta := \gamma^2 \left(v t - z\right)^2 + x_T^2$ and $\varkappa := \left(\sigma_E v \gamma/2\right) \left[\gamma \left(vt - z\right) - \sqrt{\Delta}\right]$. 
The numerical integration over all charged particles is explained in detail in Ref.\ \cite{Tuchin:2013apa}. 
With the average nuclear charge density $\bar{\rho} = Z/(\frac{4}{3} \pi R^3)$,  the magnetic fields of the projectile nucleus moving in $+z$-direction and the one of the target nucleus moving in $-z$-direction are evaluated as:
\begin{widetext}
 \begin{eqnarray}
   \boldsymbol{B}_{\text{proj}}\!\left(x_-, \vert \boldsymbol{b}_1\vert \right) &=& 
     \int \mathrm{d}^2 b^{\prime}  \, 2\, \bar{\rho} \, \sqrt{R^2 - {b^{\prime\,2}}} \,\, \boldsymbol{B}\!\left(x_-, \vert \boldsymbol{b}_1 - \boldsymbol{b}^{\prime} \vert\right) \cdot 
     \left(-\sin\psi_1 \boldsymbol{e}_x + \cos\psi_1 \boldsymbol{e}_y\right) 
     \;, \\
   \boldsymbol{B}_{\text{targ}}\!\left(x_+, \vert \boldsymbol{b}_2\vert \right) &=& 
     \int \mathrm{d}^2 b^{\prime}  \, 2\, \bar{\rho} \, \sqrt{R^2 - {b^{\prime\,2}}} \,\, \boldsymbol{B}\!\left(x_+, \vert \boldsymbol{b}_2 - \boldsymbol{b}^{\prime} \vert\right) \cdot 
     \left(-\sin \psi_2 \boldsymbol{e}_x + \cos\psi_2 \boldsymbol{e}_y\right) 
     \;,
 \end{eqnarray}
\end{widetext}
where $x_{\mp} := t \mp z/v$, while $\psi_{1,2}$ is the angle between the vector $\boldsymbol{b}_{1,2} - \boldsymbol{b}^{\,\prime}$ and the $x$-axis, 
 \begin{equation}
   \cos \psi_{1,2} = 
     \dfrac{b_{1,2} \cos \phi_{1,2} - b^{\prime} \cos\phi^{\prime}}
     {\sqrt{b_{1,2}^2 + b^{\prime \, 2} - 2  \, b_{1,2} \, b^{\prime} \cos\!\left(\phi^{\prime} - \phi_{1,2}\right)}}  \;.
 \end{equation}
We note that the total magnetic field is then simply the sum of the two, $\boldsymbol{B}_{\text{tot}} = \boldsymbol{B}_{\text{proj}} + \boldsymbol{B}_{\text{targ}}$, and while the transformation from cylindrical to Cartesian coordinates is straightforward, the transformation from Cartesian to Milne coordinates reads as~\cite{Inghirami:2019mkc}:
  \begin{equation}
    \tilde{B}_x = \dfrac{B_x}{\cosh \eta_S } \;,
  \quad
    \tilde{B}_y = \dfrac{B_y}{\cosh \eta_S} \;,
  \quad
    B_{\eta_S}  = \dfrac{B_z}{\tau} \;.
  \end{equation} 
 The initial magnetic field is shown in Fig.\ \ref{Fig:InitMagneticField1} for a Au-Au collision at $\sqrt{s_{\text{NN}}} = 200$ GeV and an impact parameter of $b=10$ fm for a set-up with $\sigma_E = 5.80$ MeV and $\sigma_\chi =0$ (top row) and a set-up with $\sigma_E = 5.80$ MeV and $\sigma_\chi =1.50$ MeV (bottom row).
  Clearly, the symmetry under reflection at the $y-$axis is broken in case of a nonvanishing value of $\sigma_\chi$, which is a consequence of the current that is produced due to the chiral imbalance. 
  This current in turn generates the symmetry-breaking magnetic field.
  It is worthwhile to remark that, although we initialize the electromagnetic fields by assuming a finite conductivity, during the solution of the hydrodynamic equations coupled with electromagnetic fields, we use the ideal-MHD approximation, where the conductivity is actually assumed to be infinite. 
  The resistive RMHD equations and the relativistic dissipative resistive RMHD equations have recently been studied in Refs.~\cite{Most:2021rhr, Nakamura:2022ssn, Dash:2022xkz}. 
   \\
%\FloatBarrier
%
% % % % % % % % % % % % % % % % % % % % % % % % % % % % % % % % % % % % % % % %
% COOPER-FRYE                                                                 %
% % % % % % % % % % % % % % % % % % % % % % % % % % % % % % % % % % % % % % % %
\subsection{Cooper-Frye prescription} 
The fireball produced in a relativistic heavy-ion collision will approach local thermodynamical equilibrium through multiple interactions of the constituent particles. 
Consequently, a thermodynamical-equilibrium pressure with respective gradients is established, which drives the expansion of the fluid into the surrounding vacuum, accompanied by cooling.
Once the temperature falls below the transition temperature between the QGP and the hadronic phase, the system hadronizes. 
Upon further cooling, the hadrons cease to interact and propagate as free-streaming particles.
The freeze-out of particles from the fluid is commonly accomplished using the widely adopted Cooper-Frye prescription \cite{Cooper:1974mv}. 
The Cooper-Frye method essentially counts the number of particle worldlines passing through the freeze-out hypersurface, which is commonly taken to be a surface of constant temperature or energy density. 
The freeze-out hypersurface is a three-dimensional surface $\Sigma\!\left(x\right)$ in four-dimensional space-time. 
This hypersurface is subdivided into infinitesimal elements $\mathrm{d} \sigma$, each associated with an outward-pointing four-vector $\mathrm{d} \Sigma_{\mu}\!\left(x\right)$ perpendicular to $\Sigma$ at point $x$.
The number of particle worldlines passing through this surface is determined by the expression:
 \begin{eqnarray}
   N_i &= &\dfrac{g_i}{\left(2 \pi \hbar\right)^3} 
     \int\limits_{\Sigma} \! \int \dfrac{\mathrm{d}^3p}{p_i^0} \, f_i\!\left(x,p_i\right) p_i^{\mu} \mathrm{d}\Sigma_{\mu}  
     \;, \label{CooperFrye1}
 \end{eqnarray}
where $g_i$ is the degeneracy factor that accounts for the internal number of degrees of freedom (i.e., spin, isospin, etc.) of particles with four-momentum $p_i^\mu := (p_i^0, \boldsymbol{p})^T$ and chemical potential $\mu_i$, while $p_i^0 \equiv E_i := \sqrt{\boldsymbol{p}^2 + m_i^2}$ is the on-shell energy of these particles, which have a mass $m_i$. 
If the fireball is in local thermodynamical equilibrium, the Lorentz-invariant local-equilibrium distribution $f_i\!\left(x,p_i\right)$ is given by:
 \begin{equation}
   f_i\!\left(x,p_i\right) = 
     \left(\exp\left\{\beta(x) \left[-p_i^{\nu} u_{\nu}(x) - \mu_i(x)\right]\right\} + a_i \right)^{-1}
     \;. \label{CooperFrye2}
 \end{equation}
Here, $a_i$ accounts for the proper quantum statistics of particles species $i$: $a_i = -1$ for bosons, $a_i = +1$ for fermions, and $a_i = 0$ for Boltzmann particles. 
The factor $-p_i^{\nu} u_{\nu}\!\left(x\right)$, where $u^{\nu}(x)$ is the fluid four-velocity, represents the energy of the particle in the local rest frame (the sign accounts for our choice of metric), while $\beta(x) := 1/T(x)$ is the inverse temperature. 
The fluid four-velocity, the temperature, and the chemical potential on the freeze-out hypersurface are obtained from the hydrodynamical calculation. 
Taking the differential of Eq.\ (\ref{CooperFrye1}) with respect to the Lorentz-invariant momentum-space measure $\mathrm{d}^3p/p_i^0$ yields the Cooper-Frye formula for particle species $i$:
 \begin{equation}
   E_i \, \dfrac{\mathrm{d}N_i}{\mathrm{d}^3p} = \dfrac{g_i}{\left(2 \pi \hbar\right)^3} \int\limits_{\Sigma} f_i\!\left(x,p_i\right) p_i^{\mu} \mathrm{d}\Sigma_{\mu} \;. \label{CooperFrye3}
 \end{equation}
Evaluating Eq.\ (\ref{CooperFrye3}) requires finding the location of the hypersurface $\Sigma$ and its normal on a discrete grid, which can be achieved by a simple comparison: 
A cell $i$ contains part of the hypersurface if its temperature $T_i$ fulfills the following inequality:
 \begin{equation}
   \left(T_{i+1} - T_{f}\right) \left(T_{f} - T_i\right) \ge 0 \;.\label{CooperFrye4}
 \end{equation}
Here, $T_{f}$ is the freeze-out value of the temperature, while $T_{i+1}$ is the temperature in one of the neighboring cells. 
Once we have found the hypersurface, we employ the \texttt{CORNELIUS} method \cite{Huovinen:2012is} to calculate the discretized normal vector $\Delta \Sigma_{\mu}$ of a two- or three-dimensional isosurface element in a three- or four-dimensional discrete grid. 
This then allows to evaluate a discretized version of the Cooper-Frye formula (\ref{CooperFrye3}), namely
 \begin{equation}
   E_i \, \dfrac{\mathrm{d} N_i}{\mathrm{d}^3p} \approx 
   \dfrac{g_i}{\left(2 \pi \hbar\right)^3} \sum_{\sigma} \Delta \Sigma_{\mu} p^{\mu} f_i\!\left(x,p_i\right)  
   \;. \label{CooperFrye5}
 \end{equation}
The four-momenta of the outgoing particles can then be obtained by a rejection sampling of Eq.\ (\ref{CooperFrye5}).

However, since in this work we do not consider the final hadronic interactions after the hydrodynamic phase \cite{Garcia-Montero:2021haa,Gotz:2021dco,Steinheimer:2017vju}, we just calculate thermal spectra based on the masses and the degeneracies of the particles, thus avoiding the noise of statistical fluctuations and saving computational time.
This approximation is legitimate, since in this work we do not yet aim at a comparison to experimental data.
%

%\FloatBarrier
% % % % % % % % % % % % % % % % % % % % % % % % % % % % % % % % % % % % % % % %
% ANISOTROPIC FLOW                                                            %
% % % % % % % % % % % % % % % % % % % % % % % % % % % % % % % % % % % % % % % %
 \subsection{Anisotropic flow}
Pressure gradients inside the fireball will produce a collective flow of particles, resulting in a radially symmetric expansion if the fireball is azimuthally symmetric in the transverse plane. 
However, in non-central collisions, the fireball takes on an almond-like shape due to the nonvanishing impact parameter, cf.\ Fig.\ \ref{Fig:InitEnergyDensity1}, which will lead to stronger pressure gradients in the $x$-direction (i.e., in the reaction plane) than in the $y$-direction (i.e., perpendicular to the reaction plane), leading to an anisotropic flow of matter, which is stronger in $x$- than in $y$-direction. 
This anisotropic flow is therefore an observable that reflects the geometry of the collision zone, describing the correlation between the direction of the flow and the distribution of matter. 
The anisotropic flow is reflected in the momentum distribution of the particles emitted from the system. 
In order to quantify this, one decomposes the azimuthal momentum distribution into a Fourier series,
  \begin{equation}
   \dfrac{\mathrm{d} N_i}{p_T \, \mathrm{d}p_T \mathrm{d}Y_i \mathrm{d}\phi} = 
   \dfrac{\mathrm{d} N_i}{2 \pi \, p_T \, \mathrm{d}p_T \mathrm{d}Y_i} \left[1 + 2 \sum\limits_{n=1}^{\infty} v_n \cos\!\left(n  \phi\right)\right]  
   \;, \label{AnisoFlow1}
  \end{equation}
where $p_T := \sqrt{p_x^2 + p_y^2}$ is the transverse momentum, $Y_i := \frac{1}{2} \ln \left[ (E_i +p^z)/(E_i-p^z)\right]$ is the  longitudinal rapidity of hadron $i$, and the Fourier coefficient $v_n$ of the $n$-th harmonic is defined as:
  \begin{equation}
    v_n := \langle \cos\!\left(n \phi\right) \rangle  \;. \label{AnisoFlow2}
  \end{equation}
Here, $\langle ... \rangle$ denotes the average over azimuthal angle, i.e.,
  \begin{equation}
    v_n = \dfrac{\int\limits_0^{2 \pi} 
      \dfrac{\mathrm{d} N_i}{p_T \, \mathrm{d}p_T \mathrm{d}Y_i \mathrm{d}\phi} \, 
      \cos\!\left(n \phi\right) \mathrm{d}\phi}{\int\limits_0^{2 \pi} \dfrac{\mathrm{d} N_i}{p_T \, \mathrm{d}p_T \mathrm{d}Y_i \mathrm{d}\phi} \mathrm{d}\phi}  \;. \label{AnisoFlow3}
  \end{equation}
The first of these harmonics, $v_1$, is referred to as directed flow. 
It describes the deflection of particles away from the beam axis, primarily caused by the bounce-off effect from the fireball matter. 
Hence, $v_1$ serves as a measure for quantifying the total amount of transverse flow.
The far more prominent observable is the second coefficient, $v_2$, or elliptic flow, whose definition is
  \begin{equation}
    v_2 = \langle \cos\!\left(2 \phi\right) \rangle = \left\langle \dfrac{p_x^2 - p_y^2}{p_x^2 + p_y^2} \right\rangle  
    \;, \label{AnisoFlow4}
  \end{equation}
and which is most closely related to the initial anisotropy of the colliding nuclei. 
As explained above, the initial energy-density distribution exhibits azimuthal asymmetry in the transverse plane, leading to stronger pressure gradients along the $x$-direction compared to the $y$-direction. 
Consequently, the fluid will have a higher velocity in the $x$-direction than in $y$-direction, resulting in a non-zero and positive value for $v_2$.
Therefore, the elliptic flow is a useful observable for gaining an understanding of the initial anisotropy of the fireball matter. 
% % % % % % % % % % % % % % % % % % % % % % % % % % % % % % % % % % % % % % % %
% RESULTS                                                                     %
% % % % % % % % % % % % % % % % % % % % % % % % % % % % % % % % % % % % % % % %
\section{Results of Numerical Simulations for Heavy-Ion Collisions}
In this section, we present some results of relativistic Au-Au collisions calculated with \texttt{BHAC-QGP}. 
Since we are particularly interested in using \texttt{BHAC-QGP} to study the effects of magnetic fields on the QGP, we first start with an investigation of the lifetime of the magnetic field, comparing the numerical results with those of \texttt{ECHO-QGP} \cite{Inghirami:2016iru}.
Finally, we showcase particle spectra for different Au-Au collisions at RHIC energies, specifically at $\sqrt{s_{\text{NN}}} = 200~\text{GeV}$. 
These results are not meant to reproduce experimental results, but are intended to highlight the diverse capabilities of \texttt{BHAC-QGP}. 
\par
If not stated otherwise, we adopt a constant inelastistic nucleus-nucleus cross section of 
$\sigma_{\text{inel}} = 4.2~\text{fm}^2$ and Milne coordinates, which are defined as
  \begin{equation}
    \left(\tau, x, y, \eta_S\right) := \left(\sqrt{t^2 - z^2}, x, y, 
    \dfrac{1}{2} \text{ln}\left(\dfrac{t + z}{t - z}\right)\right)  \;. \label{Results1}
  \end{equation}
We assume that the fireball approaches local thermodynamical equilibrium rather fast and set as initial time $\tau_0 = 0.4~\text{fm}$. 
The system of hydrodynamic equations is closed with the equation of state for a massless gas of classical particles, $P = e/3$. 
For the computation of the particle spectra, we assume that the particle density $n$ is given by the equation $n = g ~T^3 ~ /~ [\pi^2 \left(\hbar c\right)^3]  $, with $g$ = 37.
For the (2+1)-dimensional simulations we define a computational domain of $[-15~\mathrm{fm}, +15~\mathrm{fm}]^2$, with 150 cells in each direction, while the (3+1)-dimensional calculations are performed on a grid with size $[-20~\mathrm{fm}, +20~\mathrm{fm}]^2 \times [-20,20]$ and 200 cells in each direction. 
For the initialization of the energy density we set $e_0 = 55~\text{GeV}/\text{fm}^3$, while the longitudinal expansion is computed with $\Delta_{\eta_S} = 5.9$ and $\sigma_{\eta_S} = 0.4$. 

\begin{figure*}[!htp]
   \centering
    \includegraphics[width=0.98\textwidth]{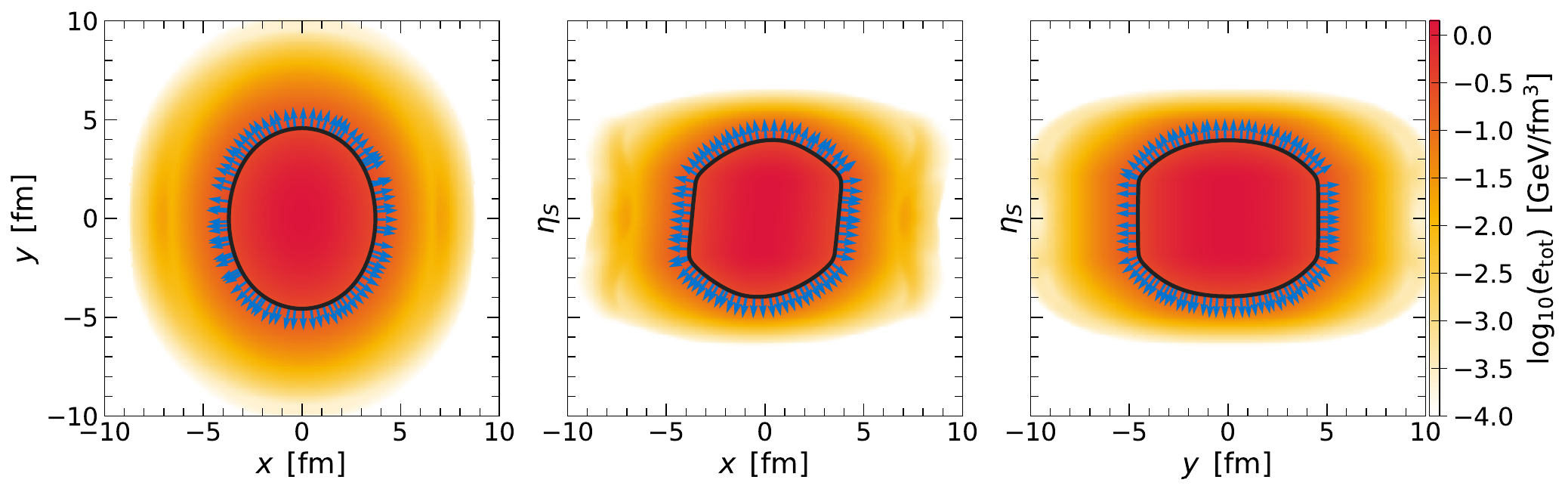}
    \caption{Total-energy density distribution at $\tau = 3.4~\text{fm}$ ($\tau_0 = 0.4~\text{fm}$) resulting from a Au-Au collision at $\sqrt{s_{\text{NN}}} = 200~\text{GeV}$, and an impact parameter of $b = 8~\text{fm}$. 
    }
\label{Fig:e-density_tau=3.4fm}
  \end{figure*}

In Fig.\ \ref{Fig:e-density_tau=3.4fm} we show the energy-density distribution for a Au-Au collision at $\sqrt{s_{\text{NN}}} = 200~\text{GeV}$ at an impact parameter of $b = 8~\text{fm}$ as computed by \texttt{BHAC-QGP} at $\tau = 3.4$ fm for an initial time $\tau_0 = 0.4~\text{fm}$.
The contour of the freeze-out energy-density $e_{f} = 0.3~\text{GeV}/\text{fm}^3$ is indicated by the black line. 
The blue arrows correspond to the normal vectors of the hypersurface, which are evaluated with the \texttt{CORNELIUS} method \cite{Huovinen:2012is}. 
For visibility, only every fourth
normal vector is plotted.

  % % % % % % % % % % % % % % % % % % % % % % % % % % % % % % % % % % % % % % % %
%                                                                             %
% % % % % % % % % % % % % % % % % % % % % % % % % % % % % % % % % % % % % % % %
  \subsection{Decay of magnetic field}
\begin{figure}[!ht]
   \centering
   \includegraphics[width=0.95\linewidth]{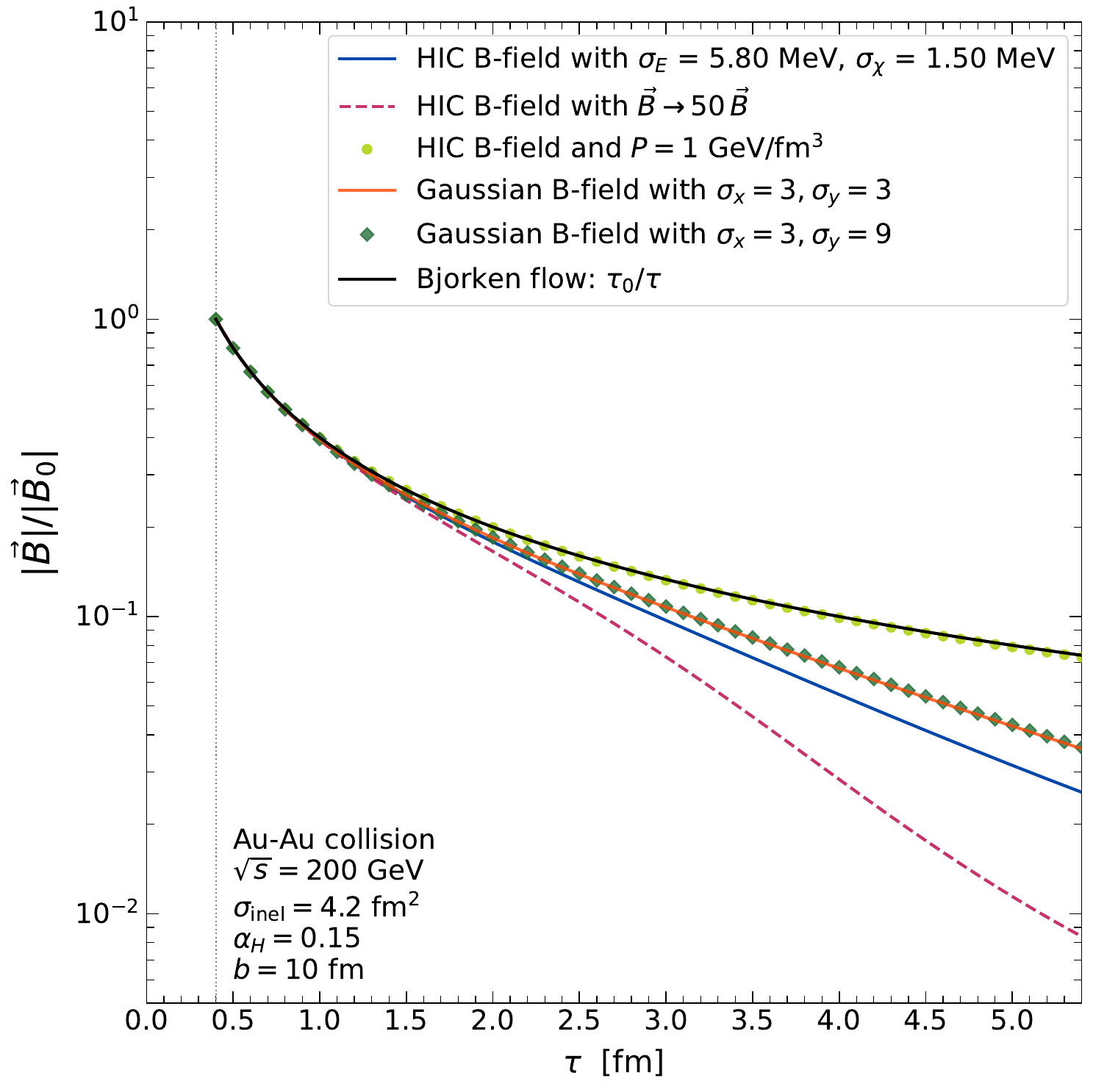}
   \caption{Time evolution of the strength of the magnetic field at the center of the grid 
   $(x = y = 0~\mathrm{fm})$. 
   Various magnetic-field configurations are compared with the magnetic field of a heavy-ion collision (HIC) and the one of Bjorken's solution.    
   The magnetic field of the HIC in \texttt{BHAC-QGP} is initialized with an electric conductivity of 
   $\sigma_E = 5.8~\text{MeV}$ and a chiral conductivity of $\sigma_\chi = 1.5~\text{MeV}$.
   }
   \label{Fig:DeacayMagneticField1}
 \end{figure}

Since the lifetime of the magnetic field is a major uncertainty in heavy-ion collisions, we investigate the temporal evolution of the magnetic field strength squared $\boldsymbol{B}^{2}$ for different cases. 
We compare the decay of the magnetic field strength at the center of the grid with that for generalized Bjorken flow and various other field configurations. 
The reference magnetic field corresponds to that of a peripheral Au-Au collision $(b=10~\text{fm})$ at RHIC energies of $\sqrt{s_{\text{NN}}} = 200~\text{GeV}$, while the energy density is initialized according to the optical Glauber model, see Eq.\ (\ref{GlauberEq9}), with $\alpha_H = 0.15$ and $\sigma_{\text{inel}} = 4.2~\text{fm}^2$. 
For the QGP, we assume an electric conductivity of $\sigma_E = 5.8~\mathrm{MeV}$ and a chiral conductivity of $\sigma_{\chi} = 1.5~\mathrm{MeV}$.
Both the fluid and the magnetic field contribute to the total pressure, $P_{\text{tot}} = P + B^2/2$, causing the matter to expand. 
In Fig.\ \ref{Fig:DeacayMagneticField1}, we compare the evolution of this magnetic field (solid blue line in the figure) with that for generalized Bjorken flow, corresponding to the evolution of a uniform pressure and magnetic field. 
As Refs.\ \cite{Pu:2016ayh, Roy:2015kma} have shown, in such a scenario the magnetic field evolves according to $B/B_0 = \tau_0/\tau$ (solid black line in the figure). 
In contrast, the magnetic field of a heavy-ion collision is not uniform, but forms additional pressure gradients, which are responsible for a faster decay. 
The stronger the magnetic field is, the stronger the pressure gradients that are formed, which drive matter faster away from the center (see dashed red line in the figure for which each component of the magnetic field was enhanced by a factor of 50).
As an additional comparison, we have also explored the case where the fluid pressure is uniform $(P=1~\mathrm{GeV}/~\mathrm{fm}^3)$, while the magnetic field mirrors that of a heavy-ion collision (green filled circles in Fig.\ \ref{Fig:DeacayMagneticField1}). 
In this case, the evolution of the magnetic field is more similar to the one of generalized Bjorken flow, with small differences, resulting from the magnetic-pressure gradients that are absent for generalized Bjorken flow.
Also in this case, increasing the magnetic field would cause a faster time decay.
%Similar results are obtained if the magnetic field is set to a constant value of $B_x = B_y = 1/\sqrt{50}~\mathrm{GeV}^{1/2}/\mathrm{fm}^{3/2}$, but the fluid pressure is initialized according to the Glauber model. 
%In this case, the pressure gradients are driving the matter away from the center, resulting in a faster decay of the magnetic field strength, when compared to generalized Bjorken flow.
If, on the other hand, the fluid pressure is set up according to the optical Glauber model and a Gaussian distribution for $B_x$ and $B_y$ is chosen, the total pressure gradients are stronger and the decay is correspondingly stronger. 
However, there is no noteworthy difference between a symmetric $(\sigma_x = 3~\mathrm{fm}, \sigma_y = 3~\mathrm{fm})$ and an asymmetric $(\sigma_x = 3~\mathrm{fm}, \sigma_y = 9~\mathrm{fm})$ distribution of the magnetic field components (solid orange line and green diamonds in the figure, respectively).
  \\
% % % % % % % % % % % % % % % % % % % % % % % % % % % % % % % % % % % % % % % %
%                                                                             %
% % % % % % % % % % % % % % % % % % % % % % % % % % % % % % % % % % % % % % % %
  \subsection{Influence of magnetic field on the QGP evolution}
The magnetic field produced in a relativistic heavy-ion collision is responsible for a pressure that, in particular outside the collision zone, is much stronger than the fluid pressure. 
Indeed, the ratio between the magnetic pressure $P_{\text{mag}} := B^2/2$ and the fluid pressure $P$, the so-called inverse plasma-$\beta$ parameter, is likely to exceed $100$ outside the collision zone. 
Such regions are very difficult to handle for MHD codes, since the conversion between conserved and primitive variables can quickly fail. 
Using the equation of state $P = e/3$, in the ideal-MHD regime \texttt{BHAC-QGP} can exploit an additional evolution equation for the entropy, as described in Paper I.
Nevertheless, to keep the initial inverse plasma-$\beta$ parameter controlled, it has proven to be advantageous to establish a relatively high minimum value for the energy density of the fluid, denoted as $e_{\text{min}}$. 
Unless otherwise specified, we use $e_{\text{min}} = 10^{-3}~\mathrm{GeV}/\mathrm{fm}^3$, which is a negligible value compared to the freeze-out energy density, $e_{f} \sim 0.5~\mathrm{GeV}/\mathrm{fm}^3$, and allows us to perform any ultrarelativistic Au-Au collision without numerical problems. 
  \begin{figure*}[!htp]
    \centering
    \includegraphics[width=0.95\textwidth]{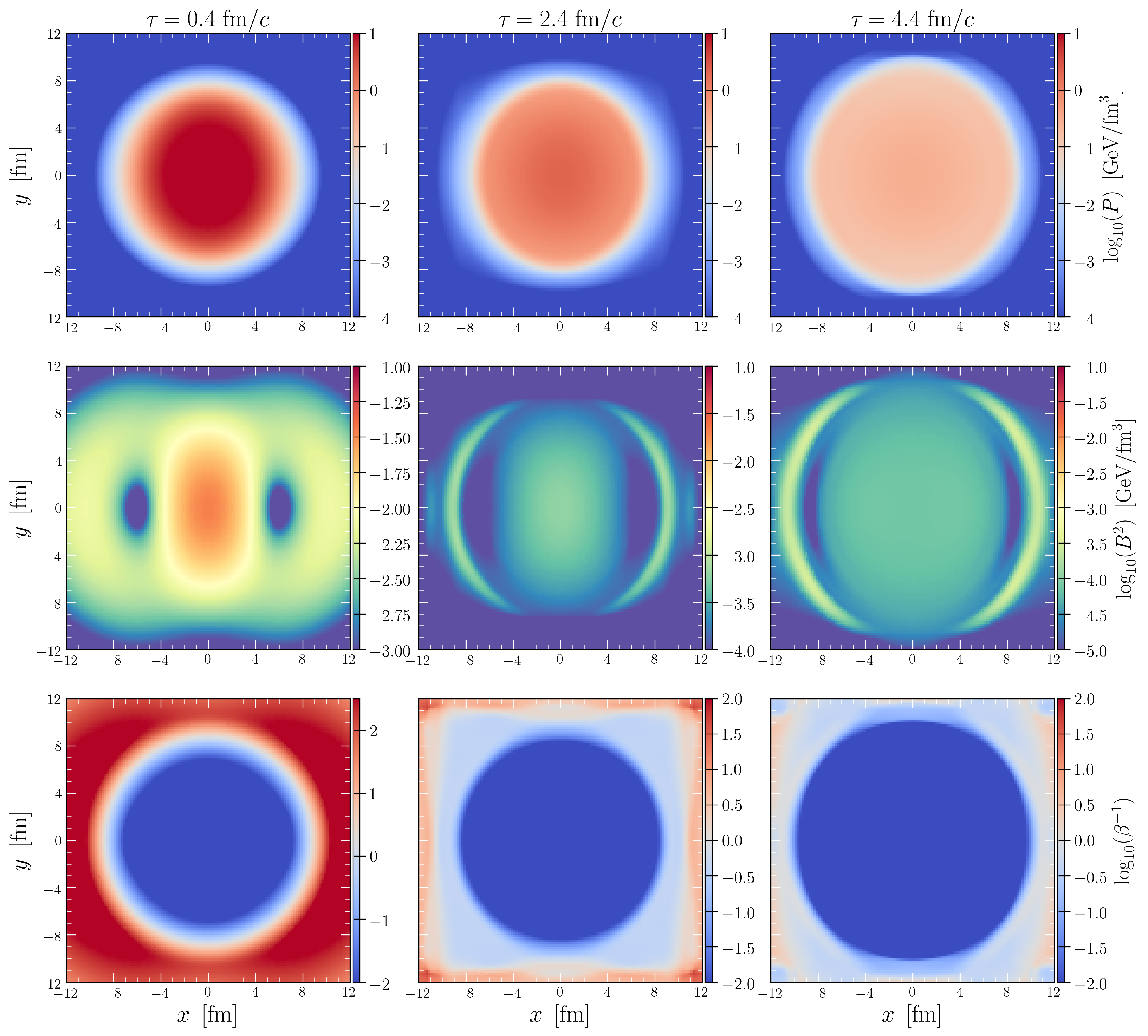}
    \caption{Evolution of the fluid pressure (top row), the magnetic field strength (second row), and the inverse plasma-$\beta$ parameter (third row) in a Au-Au collision with $b=4~\text{fm}$.
    Further parameters are listed in Tab.\ \ref{Tab:ParticleSpectra1}. 
    Since this is a rather central collision, the fluid pressure $P$ is almost isotropic and subsequently expands smoothly in all directions. 
    Compared to the fluid pressure, the magnetic field is only noticeably strong outside the fireball. 
    It has no noteworthy influence on the evolution of the fireball. 
    Albeit further away from the center some regions are highly magnetized, \texttt{BHAC-QGP} can simulate these regions with the help of the entropy evolution equation.}
    \label{Fig:HIC-Evolution1}
  \end{figure*} 
%\FloatBarrier

\begin{figure*}[!htp]
    \centering
    \includegraphics[width=0.95\textwidth]{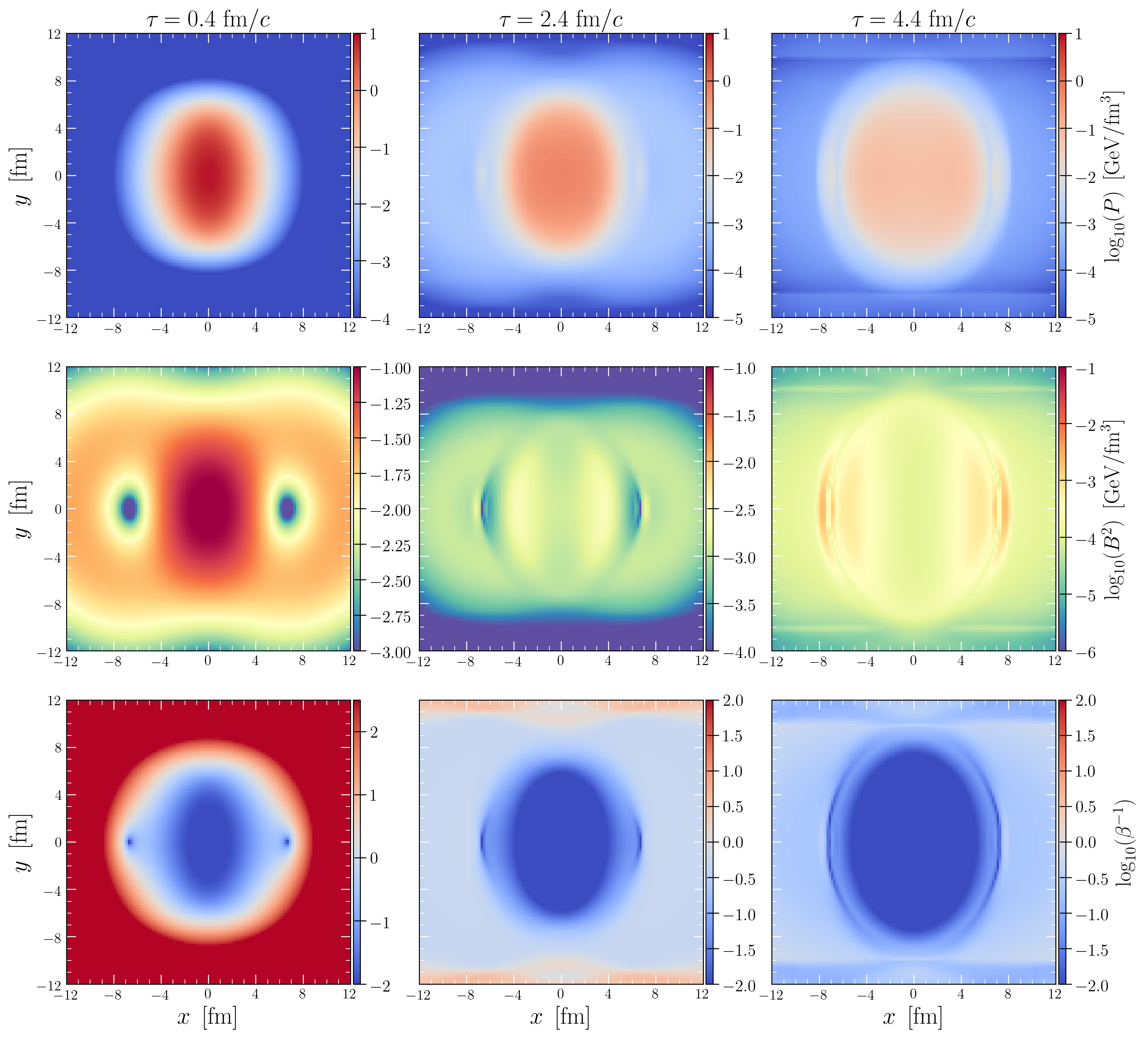}
    \caption{Evolution of the fluid pressure (top row), the magnetic field strength (second row), and the inverse plasma-$\beta$ parameter (third row) in a Au-Au collision with $b=10~\text{fm}$. 
    Further parameters are listed in Tab.\ \ref{Tab:ParticleSpectra1}. 
    Since it is a peripheral collision, the initial pressure distribution has an almond-like shape, which results in a non-isotropic expansion.
    The magnetic field is particularly strong outside the collision zone, with a region of weaker magnetic field in the vicinity of each of the original ion centers. 
    Subsequently, the magnetic field expands into these areas from all sides, resulting in interactions between them. 
    This interaction influences in particular the expansion at later times, as one can see in the plots of the third column $(\tau = 4.4~\mathrm{fm})$.
    As can be seen in the third row, the area outside the collision zone is highly magnetized, but \texttt{BHAC-QGP} is able to handle these regions using the entropy evolution equation.}
    \label{Fig:HIC-Evolution2}
  \end{figure*}

As a result of the strong magnetic pressure, the magnetic field affects the expansion of the QGP even in the ideal-MHD regime. 
To understand this influence, we compared and studied the evolution of a central collision $(b = 4~\text{fm})$ with that of a peripheral collision $(b = 10~\text{fm})$.
We show snapshots of the fluid pressure, magnetic field strength, and inverse plasma-$\beta$ parameter in Fig.\ \ref{Fig:HIC-Evolution1} (central collision) and Fig.\ \ref{Fig:HIC-Evolution2} (peripheral collision), respectively. 
In both figures, the first column refers to the initial pressure distribution of the heavy-ion collision at $\tau_0 = 0.4~\text{fm}$. 
The second column shows the respective quantities at $\tau = 2.4~\text{fm}$, while the third column refers to $\tau = 4.4~\text{fm}$. 
Clearly, the evolution of the pressure in case of the central collision is relatively uniform and smooth, as can be seen in the first row of Fig.\ \ref{Fig:HIC-Evolution1}. 
The magnetic field has no significant influence on its evolution, since the fluid pressure dominates over the magnetic pressure, especially in the collision zone. 
The total pressure gradient is then responsible for the relatively isotropic expansion of the magnetic field, which can be seen in the middle row of Fig.\ \ref{Fig:HIC-Evolution1}. 
The expanding fireball also drives the magnetic field relatively isotropically away from the center. 
The broken symmetry under reflection at the $y-$axis is a consequence of the chiral medium, since the magnetic field is dominant only far outside the collision zone. 
In fact, the magnetic pressure outside the fireball can be more than 100 times stronger than the fluid pressure, as can be seen from the third row of Fig.\ \ref{Fig:HIC-Evolution1}. 
However, as the snapshots prove, \texttt{BHAC-QGP} has no problem calculating these highly magnetized areas.
\par 
The dynamics is quite different for a peripheral collision, whose snapshots are presented in Fig.\ \ref{Fig:HIC-Evolution2}, and where the evolution of the fluid pressure is strongly influenced by the magnetic field. 
The reason for this are the two regions of lower magnetic pressure at $x = \pm 8~\text{fm}$ and $y=0~\text{fm}$. 
As a result, the magnetic field exerts a magnetic pressure on these areas, while, at the same time, the fireball expands towards these areas. 
This leads to an interaction from which shock waves emerge that will influence the evolution of the whole system. 
However, the chosen magnetic fields are not strong enough to noticeably affect the hypersurface of the energy density, at least not at the RHIC energies studied in this work. 
Note that peripheral collisions lead to large regions where $\beta^{-1} = P_{\text{mag}}/P > 10^{2}$, which \texttt{BHAC-QGP} can handle satisfactorily well.
%\FloatBarrier
\par
% % % % % % % % % % % % % % % % % % % % % % % % % % % % % % % % % % % % % % % %
%                                                 

\begin{table}[!ht]
   \begin{center}
 \begin{tabular}{|l|c|}
 \hline 
 \rule[-1ex]{0pt}{2.5ex} \textbf{Parameter} & \textbf{Value} \\ 
 \hline 
 \rule[-1ex]{0pt}{2.5ex} Initial time $\tau_0$ & $0.4~\text{fm}$ \\
 \hline 
 \rule[-1ex]{0pt}{2.5ex} Center-of-mass energy $\sqrt{s_{\text{NN}}}$ & $200~\text{GeV}$ \\
 \hline 
% \rule[-1ex]{0pt}{2.5ex} Impact parameter $b$ & $8.0~\text{fm}$ \\
% \hline 
 \rule[-1ex]{0pt}{2.5ex} Collision hardness $\alpha_H$ & $0.15$ \\
 \hline 
 \rule[-1ex]{0pt}{2.5ex} Inelastic cross section $\sigma_{\text{inel}}$ & $4.2~\text{fm}^2$ \\
 \hline 
 \rule[-1ex]{0pt}{2.5ex} Electric conductivity $\sigma_{E}$ & $5.8~\text{MeV}$ \\
 \hline 
 \rule[-1ex]{0pt}{2.5ex} Chiral conductivity $\sigma_{\chi}$ & $1.5~\text{MeV}$ \\
 \hline 
 \rule[-1ex]{0pt}{2.5ex} Width $\Delta_{\eta_S}$ & $5.9~\text{MeV}$ \\
 \hline 
 \rule[-1ex]{0pt}{2.5ex} Gaussian width $\sigma_{\eta_S}$ & $0.4~\text{MeV}$ \\
 \hline 
 \rule[-1ex]{0pt}{2.5ex} Minimum of initial energy density $e_{\text{min}}$ & $10^{-3}~\text{GeV}/\text{fm}^3$ \\
 \hline 
% \rule[-1ex]{0pt}{2.5ex} Freeze-out energy density $e_{\text{f.o.}}$ & $120~\text{GeV}/\text{fm}^3$ \\
% \hline 
 \rule[-1ex]{0pt}{2.5ex} Equation of state & $P = e/3$ \\
 \hline 
 \end{tabular}
 \caption{Parameters used for the (3+1)-dimensional simulations of Au-Au collisions at RHIC energy.}
 \label{Tab:ParticleSpectra1}
   \end{center}
 \end{table}                           %
% % % % % % % % % % % % % % % % % % % % % % % % % % % % % % % % % % % % % % % %Code Comparision
\subsection{Code Comparisons}
Given the widespread and increasing application of hydrodynamic simulations, both in relativistic heavy-ion collisions and astrophysical applications, it is crucial to estimate the systematic errors by comparing different numerical approaches and to demonstrate the general robustness of the results. 
In this section, we compare the following codes, with notes on their development history and target applications:

\begin{enumerate}
    \item \texttt{BHAC-QGP}: As noted in Paper-I, \texttt{BHAC-QGP} originates from the Black Hole Accretion Code (\texttt{BHAC}), which is a multidimensional GRMHD module for the MPI-AMRVAC framework \cite{Porth:2016rfi}. 
    Originally designed to solve the equations of general-relativistic magnetohydrodynamics (GRMHD) in arbitrary spacetimes and coordinates, the code also exploits adaptive mesh-refinement (AMR) techniques as needed. 
    The algorithm employs second-order finite-volume methods, with various schemes implemented for the treatment of the magnetic-field update on both ordinary and staggered grids.
    
    \item \texttt{ECHO-QGP}: This code has its origin in the Eulerian Conservative High-Order (\texttt{ECHO}) code, which solves classical relativistic MHD using high-order finite-difference reconstruction routines and one-wave or two-wave Riemann solvers \cite{DelZanna:2013eua, Inghirami:2016iru}.
    
    \item \texttt{VHLLE}: This is a (3+1)-dimensional relativistic-hydrodynamics code designed for simulating the expansion of the quark-gluon plasma and hadron matter in ultra-relativistic heavy-ion collisions. 
    The code uses a finite-volume method and is based on the relativistic Godunov-type   approximate Riemann solver HLLE \cite{Karpenko:2013wva}. 
    Additionally, \texttt{VHLLE} solves the equations of relativistic viscous hydrodynamics within the Israel-Stewart framework.
\end{enumerate}

For the code comparison, we choose initial conditions to represent \textcolor{blue}{a} typical Au-Au collision\textcolor{blue}{\sout{s}} at a RHIC energy of $\sqrt{s_{\text{NN}}} = 200~\text{GeV}$ with an impact parameter of $b = 8$ fm and freeze-out temperature $T_f = 135$ MeV. 
Additional parameters are summarized in Table \ref{Tab:ParticleSpectra1}. 
The electromagnetic fields are turned off for both the \texttt{BHAC-QGP} and \texttt{ECHO-QGP} codes, while shear and bulk viscosities are turned off for the \texttt{VHLLE} code. 
For \texttt{BHAC-QGP}, the number of AMR levels is set to one. 
The computational domain is defined in the range $[-20\,\text{fm}, 20\,\text{fm}]^2\times [-20,20]$, and we run the simulations at three different resolutions, namely for the following numbers of grid points: $120^3$, $200^3$, and $400^3$.

  \begin{figure*}[htp!]
    \centering
    \includegraphics[width=0.83\textwidth]{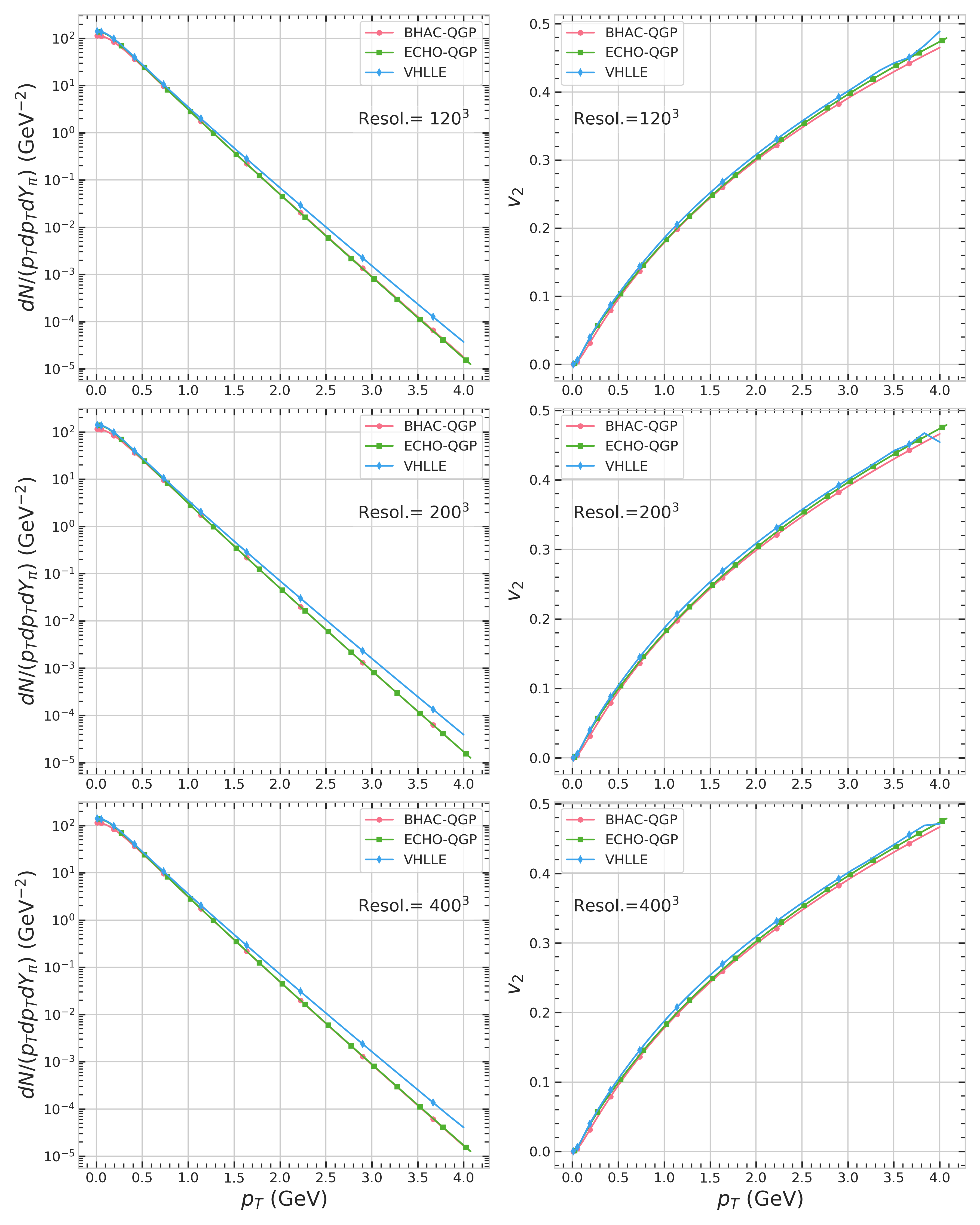}
    \caption{Comparison of particle spectra for pions (left column) and elliptic flow $v_2$ (right column) for different resolutions: top $120^3$, middle $200^3$ and buttom $400^3$. Each plot of a given rsolution also shows the results from \texttt{BHAC-QGP} red circles, \texttt{ECHO-QGP} green squares and \texttt{VHLLE} blue diamonds.}\label{fig:comparison_plot_spectra_v2_resolutions}
  \end{figure*}
In Fig.~\ref{fig:comparison_plot_spectra_v2_resolutions} we show the single-inclusive transverse-momentum spectra (left column) and the elliptic flow (right column) for pions at mid-rapidity.
The number of grid points increases from $120^3$ (top) to $200^3$ (middle) to $400^3$ (bottom). 
Each plot for a given resolution  displays the results from \texttt{BHAC-QGP} (red circles), \texttt{ECHO-QGP} (green squares), and \texttt{VHLLE} (blue diamonds).

As can be seen from the transverse-momentum spectra, at low $p_T$ \texttt{VHLLE} qualitatively agrees with both \texttt{BHAC-QGP} and \texttt{ECHO-QGP}. 
However, at high $p_T$ ($\geq 1$ GeV), \texttt{VHLLE} shows a flatter pion spectrum than both \texttt{BHAC-QGP} and \texttt{ECHO-QGP}. 
At low $p_T$, \texttt{VHLLE} produces the largest number of particles, followed by \texttt{ECHO-QGP}, and then \texttt{BHAC-QGP}. 
This difference may be attributed to a larger numerical entropy production, cf.\ discussion below.
Similarly, for the elliptic flow $v_2$, the order remains the same: \texttt{VHLLE} has the highest values, while \texttt{BHAC-QGP} shows the lowest.

We also compare each code for different resolutions, with a summary of integrated $p_T$ spectra and elliptic flow presented in Tab.~\ref{tab:comparison}. 
Interestingly, for \texttt{BHAC-QGP}, the integrated spectra decrease as the resolution increases, whereas for both \texttt{ECHO-QGP} and \texttt{VHLLE}, the integrated spectra surprisingly increase. 
Intuitively, one might expect numerical entropy to decrease with increasing resolution, leading to a reduction in integrated spectra. 
Among the codes, \texttt{VHLLE} consistently produces the highest yield at mid-rapidity across all resolutions, followed by \texttt{ECHO-QGP}, with \texttt{BHAC-QGP} yielding the lowest. 

\begin{table*}[htp!]
\centering
\begin{tabular}{|c|c|c|c|c|c|}
\hline
Code  & Resolution & Integrated Spectra/$(2\pi)$  & Integrated $v_2$ & Error Spectra (\%) & Error $v_2$ (\%) \\
\hline
\texttt{BHAC-QGP}  & 120$^3$        & 11.306612          & 2.816994         & 0.224392           & -0.042274       \\
\texttt{BHAC-QGP}  & 200$^3$        & 11.284360          & 2.816532         & 0.027147           & -0.058648       \\
\texttt{BHAC-QGP}  & 400$^3$        & 11.281298          & 2.818185         & 0.000000           & 0.000000        \\
\texttt{ECHO-QGP}  & 120$^3$        & 12.048414          & 2.857390         & -0.040782          & -0.020039       \\
\texttt{ECHO-QGP}  & 200$^3$        & 12.052316          & 2.857811         & -0.008405          & -0.005305       \\
\texttt{ECHO-QGP}  & 400$^3$        & 12.053330          & 2.857962         & 0.000000           & 0.000000        \\
\texttt{VHLLE}     & 120$^3$        & 12.744606          & 2.902196         & -0.984782          & -0.001096       \\
\texttt{VHLLE}     & 200$^3$        & 12.822257          & 2.890757         & -0.381499          & -0.395226       \\
\texttt{VHLLE}     & 400$^3$        & 12.871361          & 2.902228         & 0.000000           & 0.000000        \\
\hline
\end{tabular}
\caption{Comparison of integrated spectra, integrated $v_2$, and errors for different codes and resolutions. 
The error values are calculated with reference to the highest resolution for each respective code.}
\label{tab:comparison}
\end{table*}

%\FloatBarrier
% % % % % % % % % % % % % % % % % % % % % % % % % % % % % % % % % % % % % % % %
%                                                                             %
% % % % % % % % % % % % % % % % % % % % % % % % % % % % % % % % % % % % % % % %
 \subsection{Particle Spectra}

After presenting a comparison between the different numerical codes in the absence of a magnetic field in the previous section, we now focus on (3+1)-dimensional simulations in the presence of a magnetic field. 
In particular, we compute the particle spectra obtained from Au-Au collisions at a RHIC energy of $\sqrt{s_{\text{NN}}} = 200~\text{GeV}$, with all relevant parameters summarized in Tab.~\ref{Tab:ParticleSpectra1}. 
The impact parameter $b$ and the freeze-out temperature $T_f$ will be varied and are quoted separately.
\par
\begin{figure*}[htp!]
    \centering
    \includegraphics[width=0.9\textwidth]{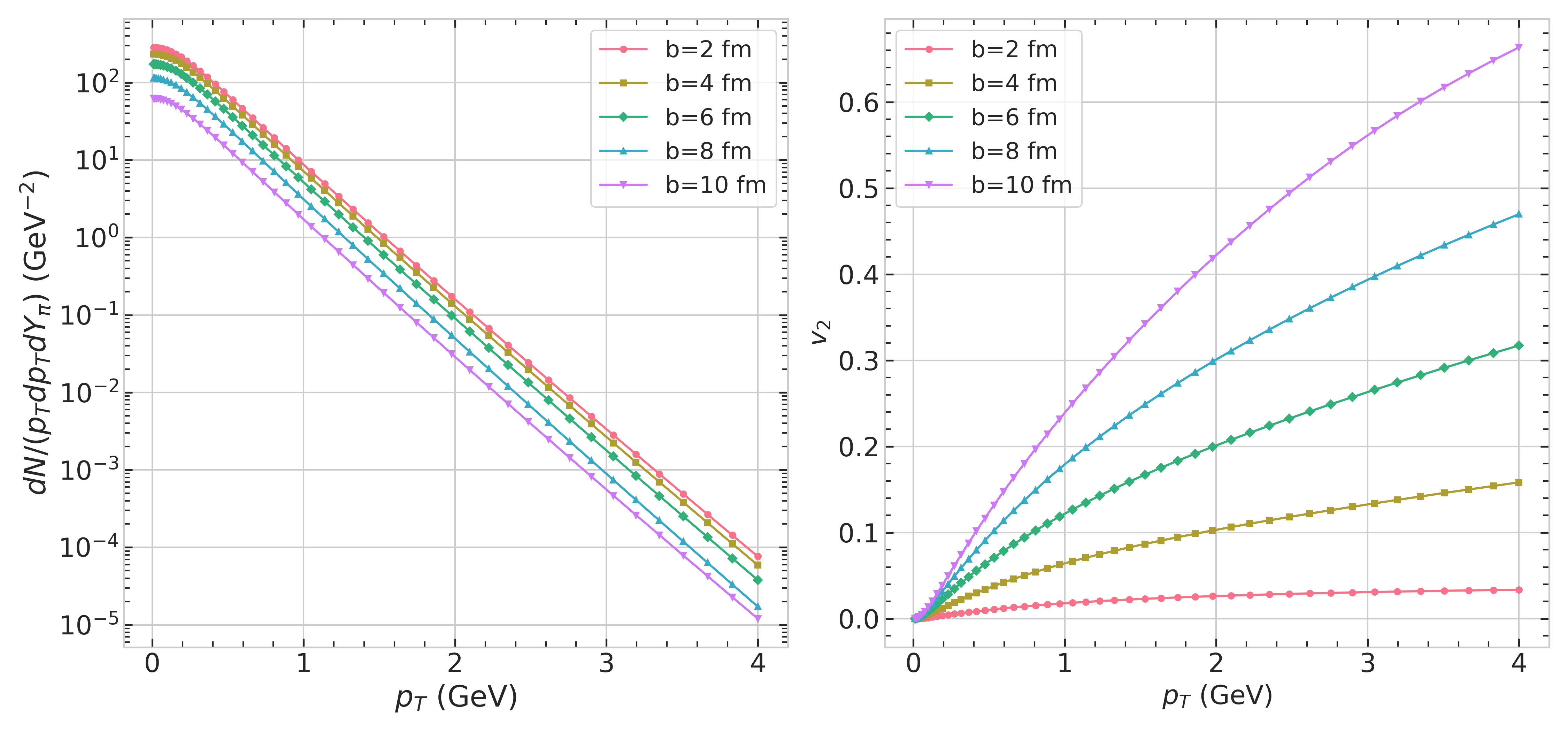}
\caption{Comparison of the transverse-momentum distribution of pions ($\pi^{\pm}, \pi^{0}$) (left panel) and elliptic flow $v_2$ as a function of transverse momentum (right panel) at mid-rapidity for Au-Au collisions, with parameters summarized in Tab.~\ref{Tab:ParticleSpectra1}, for different impact parameters at a freeze-out temperature of $T_f = 135$ MeV.
The more central the collision, the larger the fireball and the higher its energy density, leading to the generation of more particles. 
Conversely, larger impact parameters result in a larger pressure anisotropy, which produces a more pronounced elliptic flow. }\label{Fig:comparison_plot_impactParameter}
  \end{figure*}

We first vary the impact parameter $b$ while keeping the freeze-out temperature fixed at $T_f = 135~\text{MeV}$. 
Figure~\ref{Fig:comparison_plot_impactParameter} (left panel) shows the transverse-momentum spectra of pions produced at mid-rapidity for various impact parameters. 
According to the optical Glauber model, a smaller impact parameter corresponds to a larger number of participants, resulting in more energy being deposited in the collision zone. 
Consequently, collisions with smaller impact parameters produce more pions, which is clearly visible in Fig.~\ref{Fig:comparison_plot_impactParameter} (left panel). 

Additionally, more peripheral collisions produce a more pronounced anisotropy. 
In a peripheral collision, the initial fireball adopts an almond-shaped form in the transverse plane, creating larger pressure gradients in the $x$-direction compared to the $y$-direction. 
This results in a positive elliptic flow $v_2$. 
Conversely, in more central collisions, the fireball expands more uniformly, reducing the pressure anisotropy and subsequently weakening the elliptic flow. 
These trends are clearly seen in Fig.~\ref{Fig:comparison_plot_impactParameter} (right panel).
   \begin{figure*}[htp!]
   \centering
   \includegraphics[width=0.95\textwidth]{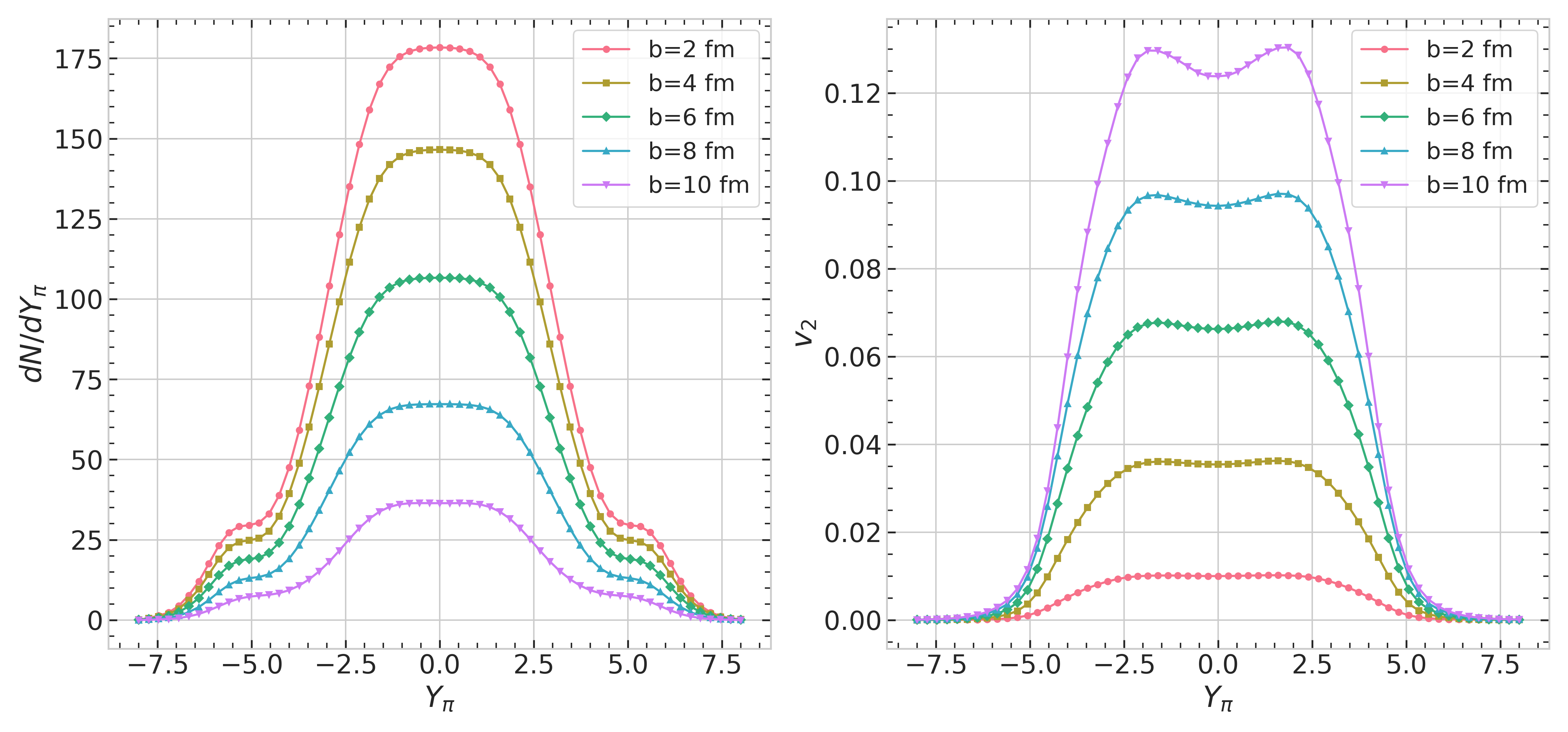}
\caption{Rapidity distribution  of  pions ($\pi^{\pm}, \pi^{0}$) (left panel) and elliptic flow $v_2$ (right panel) as a function of rapidity $Y_\pi$ for different impact parameters at a freeze-out temperature of $T_f = 135$ MeV. 
Larger impact parameters result in a stronger pressure anisotropy, leading to a more pronounced elliptic flow.
}
\label{Fig:comparison_plot_rapidity_impactParameter}
 \end{figure*}
\par
Figure~\ref{Fig:comparison_plot_rapidity_impactParameter} (left panel) shows the rapidity distribution of pions. 
Around mid-rapidity ($|Y_\pi| \leq 1$), the distribution exhibits a plateau, which is a manifestation of the Bjorken flow geometry for symmetric Au-Au collisions at high energy. 
As is well known, the final-state particle distribution is proportional to the product of the initial entropy density and transverse area, leading to an increase in particle multiplicity with decreasing impact parameter.

The effect of the initial anisotropy can also be analyzed by studying the elliptic flow of pions as a function of rapidity $Y_\pi$. 
As shown in Fig.~\ref{Fig:comparison_plot_rapidity_impactParameter} (right panel), the elliptic flow $v_2$ remains positive, a consequence of the larger pressure gradients in the $x$-direction.

  \begin{figure*}[htp!]
    \centering
    \includegraphics[width=0.9\textwidth]{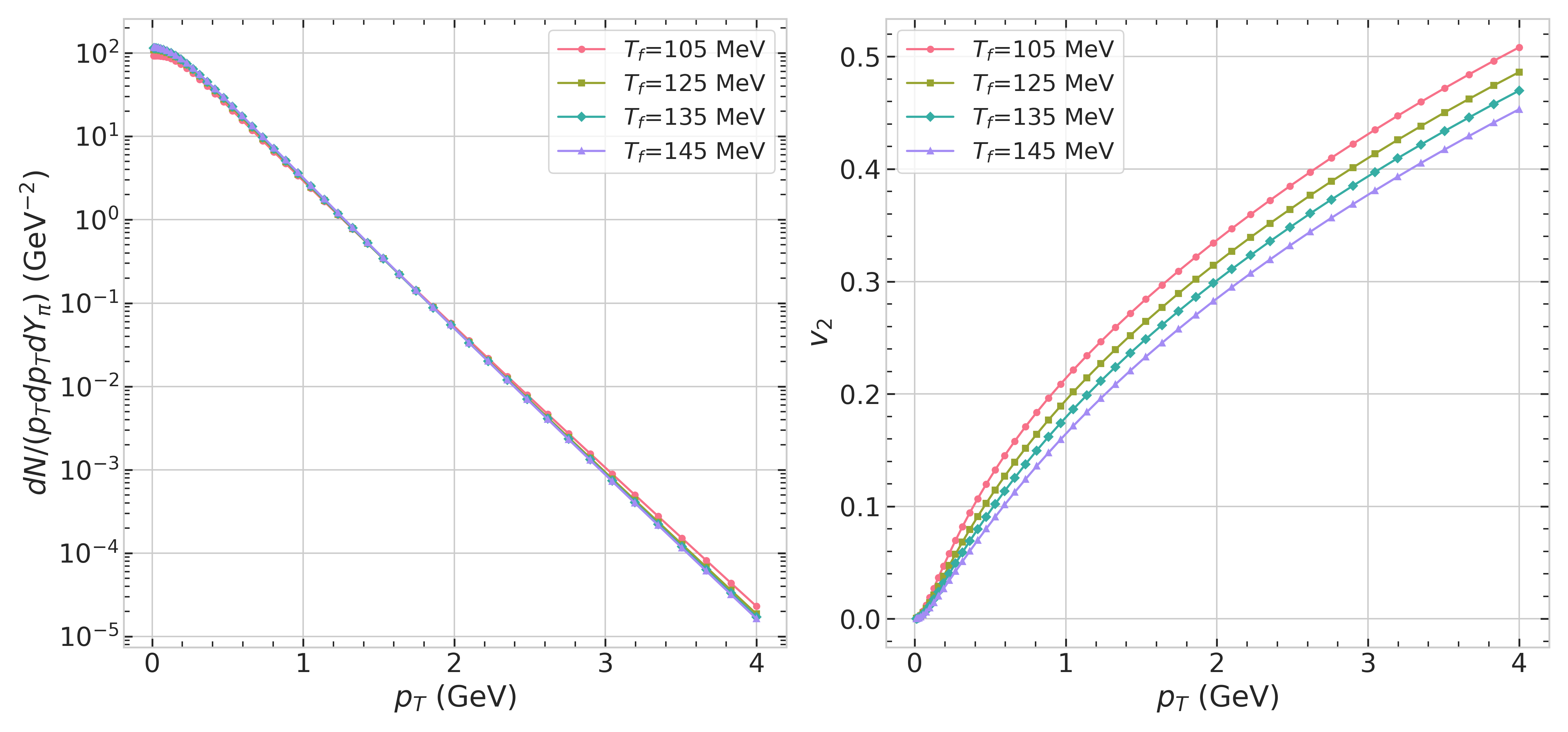}
 \caption{Comparison between the transverse-momentum distribution of pions ($\pi^{\pm}, \pi^{0}$) (left panel) and elliptic flow $v_2$ (right panel) at mid-rapidity for different freeze-out temperatures at an impact parameter of $b = 8$ fm. }\label{Fig:comparison_plot_Freezeout}
  \end{figure*}
  \begin{figure*}[htp!]
    \centering
    \includegraphics[width=0.9\textwidth]{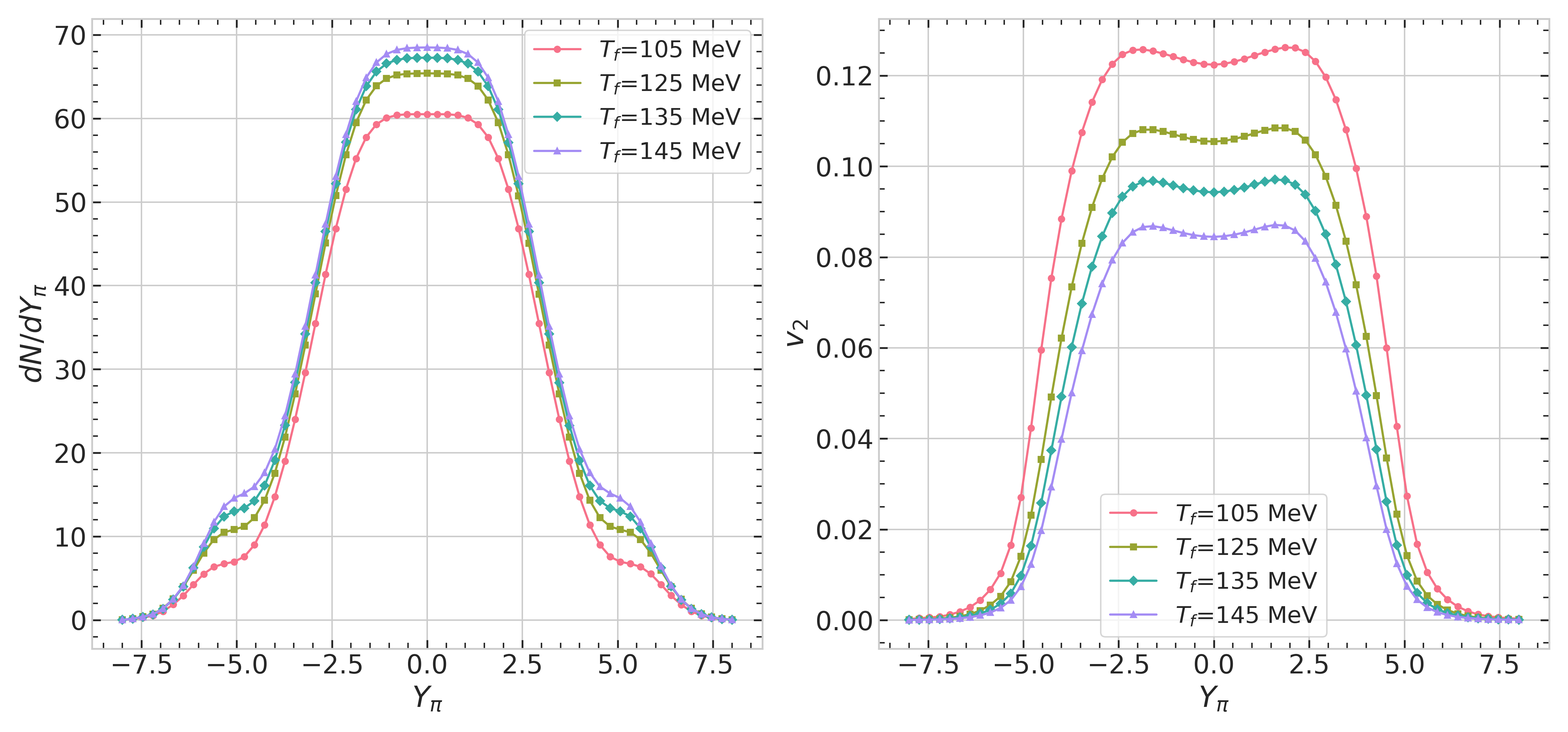}
\caption{Rapidity distribution of pions ($\pi^{\pm}, \pi^{0}$) (left panel) and elliptic flow $v_2$ (right panel) as a function of rapidity $Y_\pi$ for different freeze-out temperatures at an impact parameter of $b = 8$ fm. }
\label{Fig:comparison_plot_rapidity_Freezeout}
  \end{figure*}
 \begin{figure*}[htp!]
    \centering
    \includegraphics[width=0.9\textwidth]{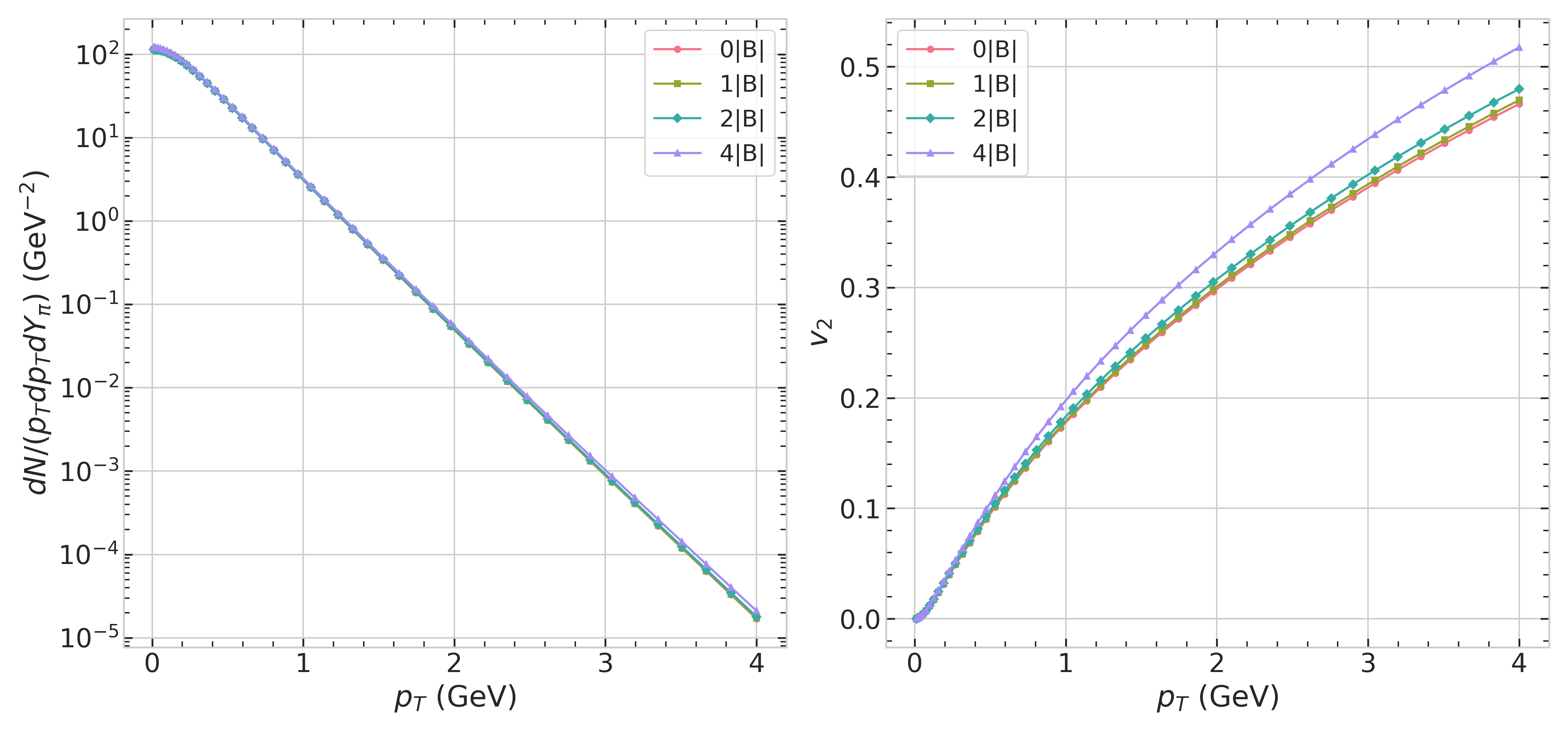}
    \caption{Effect of the magnetic field on the transverse-momentum spectrum of pions (left panel) and elliptic flow $v_2$ as a function of transverse momentum (right panel) for an impact parameter of $b = 8$ fm and a freeze-out temperature of $T_f = 135$ MeV.
    }\label{Fig:comparison_plot_diffB}
  \end{figure*}
  \begin{figure*}[htp!]
    \centering
    \includegraphics[width=0.9\textwidth]{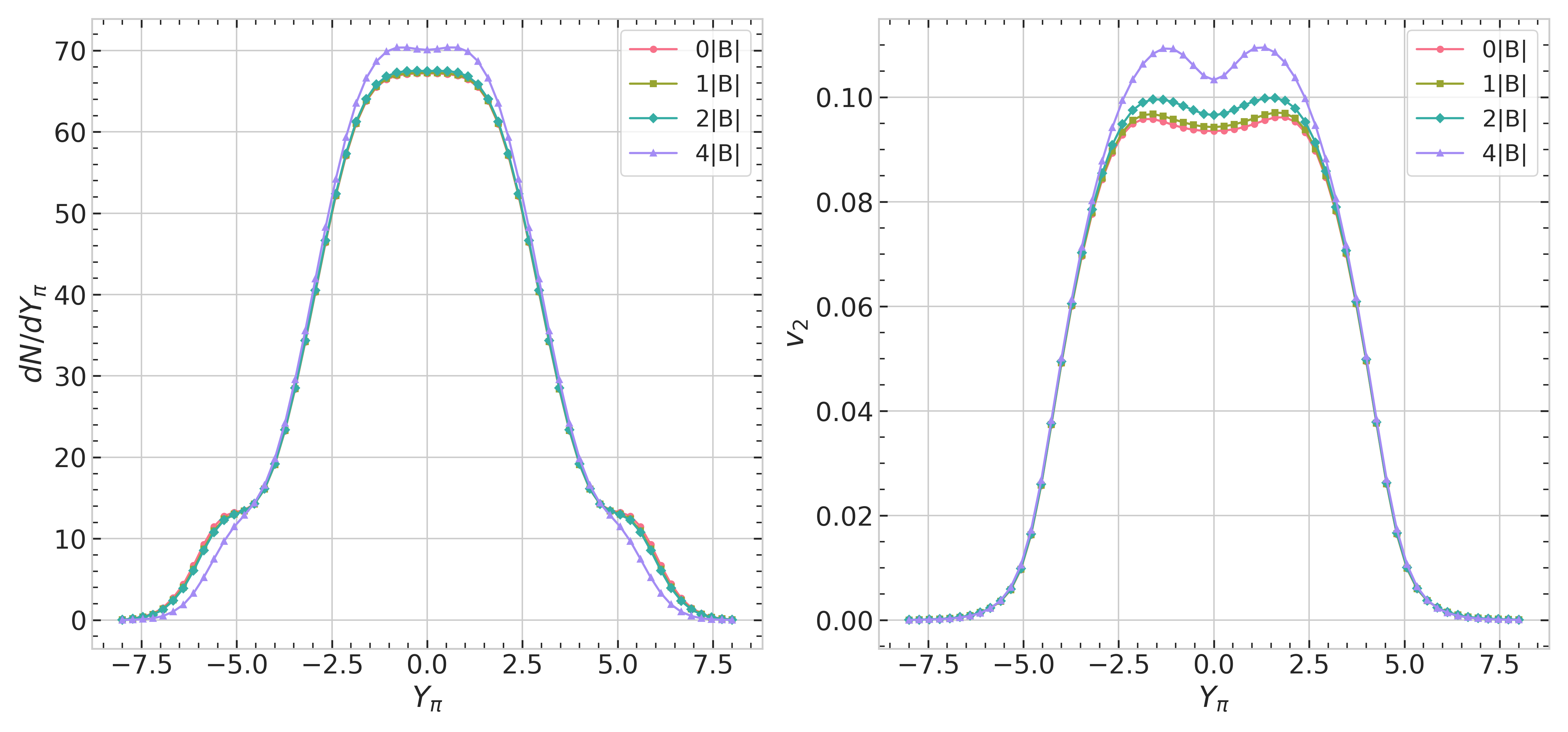}
\caption{Effect of the magnetic field on the rapidity distribution of produced pions (left panel) and elliptic flow $v_2$ (right panel) as a function of rapidity $Y_\pi$ for an impact parameter of $b = 8$ fm and a freeze-out temperature of $T_f = 135$ MeV. }\label{Fig:comparison_plot_rapidity_diffB}
  \end{figure*}

\par
We now vary the freeze-out temperature and study the effect on the number of produced pions as a function of transverse momentum and rapidity. 
Naively, we would expect a steeper slope of the transverse-momentum spectrum for a smaller freeze-out temperature.
However, this effect is counteracted by the collective flow of matter, which, for a smaller freeze-out temperature, has a longer time to build up and decreases the slope of the spectrum. 
As shown in Fig.~\ref{Fig:comparison_plot_Freezeout} (left panel), these two competing effects nearly cancel each other in the pion transverse-momentum spectrum, which thus remains approximately unaffected by changes in freeze-out temperature, except for very low $p_T$ at smaller freeze-out temperatures, where there is a slight decrease.
However, this behavior also depends on the particle species of interest. 
In general, lower freeze-out temperatures result in fewer pions, as illustrated in Fig.~\ref{Fig:comparison_plot_rapidity_Freezeout} (left panel). 
 The longer evolution time for lower freeze-out temperatures produces a more pronounced collective behavior,
which leads to an enhanced elliptic flow of pions, as seen in Fig.~\ref{Fig:comparison_plot_Freezeout} (right panel) and Fig.~\ref{Fig:comparison_plot_rapidity_Freezeout} (right panel).

\par
Investigating elliptic flow is particularly useful, as it is related to the initial anisotropy and is sensitive to the equation of state, which governs the fluid's evolution. 
To determine whether strong external magnetic fields influence the momentum anisotropy, we calculated $v_2$ for heavy-ion collisions with and without magnetic fields. 
Although the magnetic fields are strong, especially outside the fireball, they appear to have no significant effect on the energy scales relevant to heavy-ion collisions. 
This can be seen in Fig.~\ref{Fig:comparison_plot_diffB} and Fig.~\ref{Fig:comparison_plot_rapidity_diffB}, where we have plotted the results for ideal hydrodynamics without magnetic field (labelled $0|B|$ in these figures), alongside those for the standard magnetic field strength as computed in ideal RMHD (labelled $1|B|$ in these figures). 
Comparing these two curves in the figures shows that the magnetic field produced in heavy-ion collisions does not significantly impact these observables. 
This conclusion is further supported by the results obtained via \texttt{ECHO-QGP} \cite{Inghirami:2016iru}.

\par
Nevertheless, the magnetic field does influence the expansion of the fireball, especially for peripheral collisions, as discussed in Fig.~\ref{Fig:HIC-Evolution2}. 
In order to make this effect visible, we increase each component of the magnetic field by a factor of two or four,
respectively, as shown in Fig.~\ref{Fig:comparison_plot_diffB} and Fig.~\ref{Fig:comparison_plot_rapidity_diffB} (labelled $2|B|$ and $4|B|$ in these figures). 
In particular, a sufficiently strong magnetic field contributes significantly to the total energy density of the fireball, resulting in the production of more particles. 
As exemplified in the figures, when the magnetic field is increased by a factor of four, there is a substantial increase in pion production at mid-rapidity and a depletion at larger rapidities (see Fig.~\ref{Fig:comparison_plot_rapidity_diffB}, left panel). 

However, the transverse-momentum spectra for pions remain unaffected by all assumed magnetic field strengths. 
On the other hand, a higher magnetic field strength leads to stronger magnetic pressures and total pressure gradients, which result in stronger expansion and ultimately particles with higher transverse momentum.
Since these pressure gradients are also anisotropic, they further amplify the momentum anisotropy, ultimately leading to enhanced elliptic flow, as shown in Fig.~\ref{Fig:comparison_plot_diffB} (right panel) and Fig.~\ref{Fig:comparison_plot_rapidity_diffB} (right panel).
%

%
%
% % % % % % % % % % % % % % % % % % % % % % % % % % % % % % % % % % % % % % % %
%                                                                             %
% % % % % % % % % % % % % % % % % % % % % % % % % % % % % % % % % % % % % % % %
  
\section{Conclusions and Outlook}
We presented results obtained from a novel (3+1)-dimensional relativistic magnetohydrodynamics code, \texttt{BHAC-QGP}. 
\texttt{BHAC-QGP} is an extension of the Black Hole Accretion Code (\texttt{BHAC}), which is capable of evolving the matter produced in relativistic heavy-ion collisions. 
In order to perform such an evolution, we have implemented Milne coordinates and the ultrarelativistic equation of state $(P = e/3)$ for a generic investigation of heavy-ion collisions. 
We plan to implement other equations of state in the future. 
Moreover, \texttt{BHAC-QGP} can resort to the entropy evolution equation in the ideal-MHD regime, which allows to simulate also the strong magnetic fields outside the collision zone and to study novel chiral effects, such as the Chiral Magnetic Effect (CME).
\par
\texttt{BHAC-QGP} uses the optical Glauber model to initialize the energy density that is produced during 
the collision of two heavy ions. 
Therefore, all types of heavy-ion collisions with arbitrary impact parameters can be calculated. 
Additionally, our ongoing work involves implementing the option to use initial event-by-event energy-density data extracted from transport codes like \texttt{SMASH}~\cite{Weil:2016zrk}.
This will allow us to simulate a more randomly distributed and realistic energy-density profile, which is particularly crucial for the investigation of higher flow-harmonic coefficients, such as the triangular flow $v_3$. 
We expect that the AMR capability of \texttt{BHAC-QGP} will be very useful in the event-by-event study of heavy-ion collisions, as it provides the means to considerably reduce the computational effort without substantial loss of accuracy.
\par
\texttt{BHAC-QGP} relies on the \texttt{CORNELIUS} routine to evaluate the Cooper-Frye freeze-out formula. 
Subsequently, the resulting data can be used to generate particle spectra.
With this implementation, we performed a detailed code comparison and showed that \texttt{BHAC-QGP} is able to reproduce spectra from \texttt{ECHO-QGP} as well as \texttt{VHLLE}. 
\par
Another interesting topic is the investigation of a rotating QGP. 
A QGP with a non-zero angular momentum can give rise to the chiral vortical effect (CVE). 
However, its investigation requires very high resolutions. 
\texttt{BHAC-QGP}, with its AMR framework, is excellently positioned for this task. 
% % % % % % % % % % % % % % % % % % % % % % % % % % % % % % % % % % % % % % % %
%                                                                             %
% % % % % % % % % % % % % % % % % % % % % % % % % % % % % % % % % % % % % % % %
\section*{Acknowledgments}
The authors
%(s)/author(s) X (Y,Z)]
acknowledge support by the Deutsche Forschungsgemeinschaft (DFG, German Research Foundation) through the CRC-TR 211 
``Strong-interaction matter under extreme conditions'' – project number 315477589 – TRR 211.
The work is supported by the State of Hesse within the Research Cluster ELEMENTS (Project ID 500/10.006). 
Computational resources have been provided by the Center for Scientific Computing (CSC) at the Goethe University.
M.M.\ would like to thank H.\ Olivares, M.\ Chabanov, N.\ Kuebler, and J.\ Sammet for fruitful discussions.
% % % % % % % % % % % % % % % % % % % % % % % % % % % % % % % % % % % % % % % %
%                                                                             %
% % % % % % % % % % % % % % % % % % % % % % % % % % % % % % % % % % % % % % % %
%\section{References}
\newpage
\bibliography{biblio2}  % Tell bibtex which .bib file to use
%\bibliographystyle{apsrev4-2}  % Tell bibtex which bibliography style to use

%\appendix
%\section{Appendix}

\end{document}